\documentclass[authorversion]{acmart}
\usepackage{multirow}
\usepackage{makecell}
\usepackage{threeparttable}
\AtBeginDocument{%
  }

\setcopyright{acmlicensed}
\copyrightyear{2025}
\acmYear{2025}
\acmDOI{XXXXXXX.XXXXXXX}
\acmISBN{978-1-4503-XXXX-X/2025/04}

\begin{document}

\title[A Survey on Music Generation from Single-Modal, Cross-Modal and Multi-Modal Perspectives]{A Survey on Music Generation from Single-Modal, Cross-Modal and Multi-Modal Perspectives}




\author{Shuyu Li}
\affiliation{%
  \institution{College of Computer Science and Technology, Zhejiang University}
  \city{Hangzhou}
  \country{China}}
\email{lsyxary@zju.edu.cn}

\author{Shulei Ji}
\affiliation{%
  \institution{College of Computer Science and Technology, Zhejiang University, and Innovation Center of Yangtze River Delta, Zhejiang University}
  \city{Hangzhou}
  \country{China}}
\email{shuleiji@zju.edu.cn}

\author{Zihao Wang}
\author{Songruoyao Wu}
\author{Jiaxing Yu}
\affiliation{%
  \institution{College of Computer Science and Technology, Zhejiang University}
  \city{Hangzhou}
  \country{China}}
\email{{carlwang,wsry,yujx}@zju.edu.cn}



\author{Kejun Zhang}
\authornote{Corresponding Author}
\affiliation{%
  \institution{College of Computer Science and Technology, Zhejiang University, and Innovation Center of Yangtze River Delta, Zhejiang University}
  \city{Hangzhou}
  \country{China}}
\email{zhangkejun@zju.edu.cn}

\thanks{This work is supported by the National Natural Science Foundation of China  (No.62272409)}








\renewcommand{\shortauthors}{S. Li et al.}

\begin{abstract}
 Multi-modal music generation, using multiple modalities like text, images, and video alongside musical scores and audio as guidance, is an emerging research area with broad applications. This paper reviews this field, categorizing music generation systems from the perspective of modalities. The review covers modality representation, multi-modal data alignment, and their utilization to guide music generation. Current datasets and evaluation methods are also discussed. Key challenges in this area include effective multi-modal integration, large-scale comprehensive datasets, and systematic evaluation methods. Finally, an outlook on future research directions is provided, focusing on creativity, efficiency, multi-modal alignment, and evaluation.
\end{abstract}




\maketitle

\section{Introduction}
\label{sec:intro}

Humans naturally perceive and integrate multi-modal information like visual and auditory cues, but this integration remains challenging for Artificial Intelligence (AI). 
As a vital aspect of the auditory modality, music plays a significant role in human life and often accompanies text, images, and videos as background elements. Therefore, how to generate corresponding music based on multi-modal information has become one of the tasks people are currently concerned about. Large language models (LLMs) are increasingly incorporating multiple modalities to handle a wider range of tasks. 
For example, GPT-4o~\cite{openai2024gpt} integrated advanced capabilities for processing and generating visual and auditory content.
The boundaries between audio, text, images, and video are blurring. LLMs are evolving beyond single-modal information, aiming to develop a comprehensive understanding of the world. In this context, multi-modal music generation plays a crucial role. Music generation nowadays has evolved from single-modal generation to cross-modal generation and is gradually moving towards multi-modal fusion.

In the field of music generation, a considerable number of articles have reviewed the history, existing technologies, and future prospects from various perspectives. However, there is currently a lack of focus on modalities in the field of music generation. Ji et al.~\cite{deepSymbolic} reviewed the application of deep learning in symbolic music generation. Wang et al.~\cite{wang2024review} set against the backdrop of the AIGC era and reviewed advanced intelligent music generation systems.
While they touched on a few multi-modal music generation technologies, they primarily focused on symbolic music, with limited discussion of audio-based methods and comprehensive multi-modal integration. 
Dash and Agres~\cite{affectiveSurvey} reviewed AI-based affective music generation, primarily focusing on emotional guidance. 
Yu et al.~\cite{yu2024suno} reviewed the technologies, trends, and limitations in music generation after the rise of Suno AI~\footnote{\url{https://www.suno.ai}}.
Although they discussed guided music generation, they did not provide a comprehensive review of multi-modal guidance in music generation.
Ma et al.~\cite{ma2024foundation} paid attention to both symbolic music and music audio, including a review of multi-modal music generation technologies. However, this review primarily focuses on foundational models in music generation, and the discussion on multi-modality is still insufficient and not deep enough. In addition, the review categorizes existing music generation technologies according to foundational models. While it is convenient for researchers in the field of music generation to understand the widely used foundational models, it is not as convenient for researchers focusing on specific modalities. 
In the current context of music generation evolving from single-modal to cross-modal and multi-modal approaches, it is crucial to understand how different representations complement each other and reveal unique design considerations for effectively integrating diverse information sources.
Hence, a review focusing on modality interactions in music generation is indispensable.

\begin{figure} [t] 
  \centering
  \includegraphics[width=0.8\textwidth]{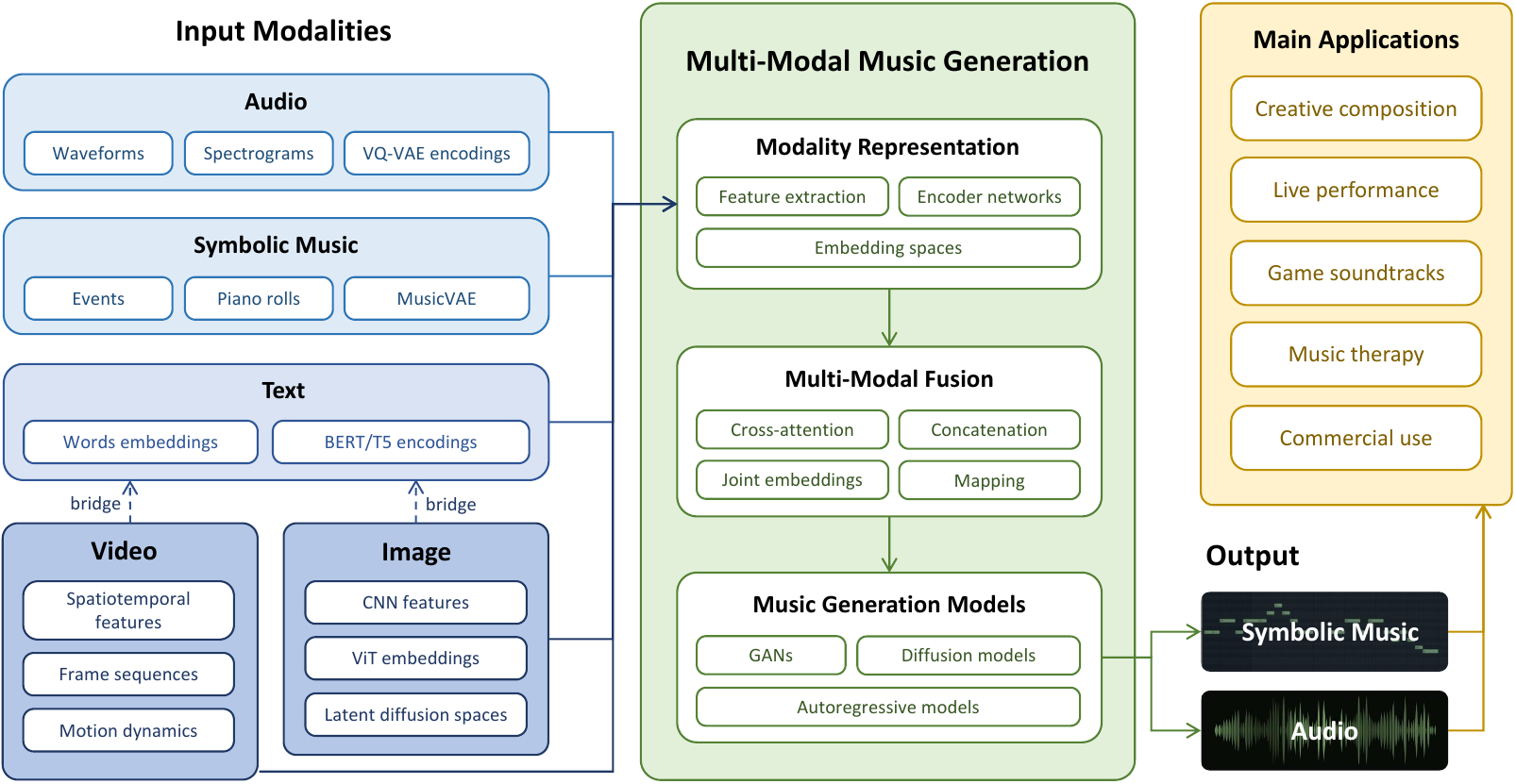}
  \caption{Overview of Multi-Modal Music Generation}
  \label{fig:overview}
  \Description{Overview}
\end{figure}

As shown in Figure~\ref{fig:overview}, multi-modal music generation methods utilize guidance modalities, leveraging multi-modal information to guide the generation of music, whether symbolic or audio. The guidance modalities include both internal musical modalities, specifically audio and symbolic music, as well as various external non-musical modalities such as text, images, and video. It is important to note that in music representation, there is a significant gap between audio and symbolic music, similar to the difference between speech and text in natural language, as their encoding methods are not the same. Therefore, in practical work, they should be treated as different modalities. Multi-modal music generation extracts features from guidance modalities through both rule-based and deep learning methods. Then the multi-modal features are integrated through methods such as cross-attention, concatenation, embedding and mapping. Also, multi-modal music generation systems can use some modalities as bridges to achieve cross-modal and multi-modal understanding. Subsequently, multi-modal music generation models generate music under the guidance of multi-modal information and achieve various downstream applications. 

Initially, music generation methods only utilized internal modalities as guidance, taking a segment of symbolic music or audio as input and generating new music through adaptation, filling, continuation, and other methods, known as single-modal music generation. Some works focus on symbolic music, achieving symbolic-to-symbolic music generation~\cite{roberts2018musicvae,wu2023melodyglm}. 
Other works use audio as guidance, achieving audio-to-audio generation~\cite{garcia2023vampnet,borsos2023audiolm}.

However, the application scope of single-modal music generation is limited. To enhance the performance and controllability of music generation and expand application scenarios, researchers have introduced external modalities as guidance modalities for music. In the early stages of this process, researchers used only a single external modality as the guidance modality for music generation, known as cross-modal music generation, such as text-to-music~\cite{copet2024musicgen,agostinelli2023musiclm,wu2023exploring}, lyrics-to-melody~\cite{ju2022telemelody,sheng2021songmass,zhang2022relyme}, and visual-to-music~\cite{sdmuse,di2021cmt,zhuo2023video} generation. 
Some works~\cite{manzelli2018conditioning,wang2019performancenet,maman2024performance} aimed to bridge the symbolic music and audio modalities, using symbolic music as the guidance modality for audio synthesis. 

With the development of multi-modal technology, researchers have begun to fuse information from multiple different modalities to jointly guide music generation, known as multi-modal music generation. 
Currently, multi-modal music generation remains in an exploratory stage. Although some successful works have been proposed~\cite{melfusion,xmusic2025,bai2024seedmusic}, the field still faces many challenges. There is still a considerable distance from the ideal comprehensive multi-modal model. 
Therefore, this article sorts out the development trajectory of music generation from the perspective of modalities, aiming to summarize experiences from this and promote research in multi-modal music generation. We will also discuss datasets and evaluation methods. Finally, we will analyze challenges and outline future research directions.

\section{Related Modalities and Representations}
\label{sec:modality}



Music generation in the multi-modal context relies on the integration of diverse modalities, each contributing unique information to guide the creative process. These modalities include audio, symbolic music, text, images, and video, each with distinct representations and characteristics. 
This section provides an overview of the key modalities and their roles in music generation, laying the foundation for the subsequent discussion of generation techniques.

\subsection{Audio}
\label{sec:audio}

Audio can be stored and processed in waveform format. However, raw waveform data are not ideal for AI models due to the high storage demands and challenges in learning stable features~\cite{wang2024review}. Consequently, audio data require compression while maintaining reconstruction quality. Researchers have long explored audio compression techniques~\cite{atal1971speech,juang1982multiple}. 

Recent advancements in deep learning have achieved superior results in audio coding models. Van Den Oord et al.~\cite{van2017vqvae} proposed VQ-VAE, introducing vector quantization to Variational Autoencoders (VAEs) for compression and discretization. Unlike VAEs’ static priors, VQ-VAE employs learnable priors, enabling flexible, high-quality compression of images, video, and audio. However, quantization errors degrade reconstruction quality. For this issue, residual vector quantization (RVQ) has proven effective~\cite{zeghidour2021soundstream}. 
Zeghidour et al.~\cite{zeghidour2021soundstream} introduced SoundStream, a CNN-based encoder-decoder architecture with RVQ. It employs adversarial training using both reconstruction and adversarial losses. Défossez et al.~\cite{defossez2022encodec} proposed EnCodec, further advancing the RVQ-based method by implementing multi-scale STFT-based discriminators and a loss-balancing mechanism to enhance perceptual quality. EnCodec also integrates Transformer-based~\cite{waswani2017attention} language models (LMs) and entropy coding for additional compression, achieving strong performance on both 24kHz mono and 48kHz stereo audio. 
Kumar et al.~\cite{kumar2024improvedrvq} proposed the Improved RVQGAN, refining the SoundStream and EnCodec framework by optimizing activation functions to eliminate jarring artifacts, redesigning codebooks to prevent collapse, and addressing side effects of quantizer dropout. 

Although adversarial methods excel in the extraction of acoustic features, they struggle with semantic understanding. In contrast, AudioMAE~\cite{huang2022audiomae} adopts a self-supervised approach inspired by MAE~\cite{he2022mae}, masking and reconstructing spectrograms to learn audio representations. Its decoder employs local window attention to capture spatiotemporal correlations, achieving state-of-the-art results in audio classification tasks without external supervision. 

Music presents unique challenges distinct from speech, requiring faithful preservation of rhythm, pitch, and timbre. Consequently, some encoding models effective in the realm of speech may not perform optimally for music encoding. For instance, the general audio encoding model EnCodec yields results inferior to those of specialized audio understanding models designed for music in MuChin benchmark~\cite{wang2024muchin}. Thus, the design and training of encoding models that address the characteristics of musical audio is pivotal for enhancing a model's capacity to process music.  Spijkervet and Burgoyne~\cite{spijkervet2021clmr} proposed the CLMR framework based on SimCLR~\cite{chen2020simclr}, constructing a music understanding model through contrastive learning. Jukebox~\cite{dhariwal2020jukebox} is a model for generating diverse, high-fidelity music within the raw audio domain. The Music VQ-VAE encoding model within Jukebox ensures reconstruction quality through a combination of sample-level reconstruction loss and STFT-based spectral loss. And this model is trained on a specialized music dataset to ensure its suitablity in music. 
Music2vec~\cite{li2022music2vec}, based on the self-supervised learning framework introduced by Data2vec~\cite{baevski2022data2vec}, is a music audio understanding model that achieves performance comparably close to that of Music VQ-VAE across various MIR downstream tasks, yet with a parameter count merely 2\% of that of Music VQ-VAE. Li et al.~\cite{li2024mert} proposed MERT, a model trained via a two-stage teacher-student framework.  MERT first learns a general acoustic representation from EnCodec~\cite{defossez2022encodec}.  Subsequently, MERT undergoes further from a music-specific teacher model to capture nuanced musical features, achieving good results. 

\subsection{Symbolic Music}


In contrast to audio, symbolic music form leans more towards the representation of musical scores. This form is adept at describing musical features such as notes and performance notations. 
Depending on the data storage methods, symbolic music can be represented in various forms such as events, piano rolls, sequences, text, and graphs~\cite{deepSymbolic}.

Since 1957, researchers have been exploring automated composition~\cite{briot2020deepSurvey}. With the continuous development of symbolic music generation research, numerous encoding methods for symbolic music have been proposed. The event-based symbolic music representation method represents music as a sequence of multiple events according to timestamps, such as note on and note off. The Musical Instrument Digital Interface (MIDI)~\footnote{\url{https://midi.org/}} is a representative of this data type and is widely used in the understanding and generation of symbolic music. However, the MIDI representation may pose difficulties for models learning musical information, such as the necessity for Note on and Note off to appear in pairs. If a model cannot understand this regularity, it cannot correctly generate music. To address the issues with MIDI in model training, Huang and Yang~\cite{remi} proposed REMI based on MIDI, resulting in a symbolic music representation more suitable for model learning. For event-based music data, the simplest processing method is to directly treat the event sequence as a 1D sequence. However, this approach often results in excessively long sequences, making it difficult for models to focus on all sequence information. Moreover, a large number of different types of events mixed in the same sequence can make it difficult for models to learn from the data. Therefore, Hsiao et al.~\cite{hsiao2021compound} classified music events into note types and metric types, grouping a series of similar events into compound words (CP), which helps models better understand musical information while shortening the sequence length. The OctupleMIDI encoding method~\cite{zeng2021musicbert} further compresses the sequence length 
compared to CP. Ren et al.~\cite{ren2020popmag} introduced a representation method suitable for multitrack symbolic music, allowing for the representation of multitrack compositions using a singular symbolic sequence. 
Roberts et al.~\cite{roberts2018musicvae} introduced a hierarchical decoder on the basis of VAEs for sequential data and developed MusicVAE, constructing a symbolic music encoding model with good reconstruction and interpolation performance. Additionally, some works represent symbolic music in formats such as piano roll representation~\cite{di2021cmt,wang2019performancenet} and text (ABC notation)~\cite{wu2023exploring}, to fully utilize different model architectures like CNNs and LMs.

With the widespread application of pre-training and self-supervised learning methods, an increasing number of works have introduced pre-training methods to achieve the encoding and generation of symbolic music. Zeng et al.~\cite{zeng2021musicbert} introduced the widely used pre-training model BERT~\cite{devlin-etal-2019-bert} from the natural language processing (NLP) field into the domain of symbolic music understanding, constructing an effective pre-training model for symbolic music through OctupleMIDI encoding and bar-level masking strategy. Sheng et al.~\cite{sheng2021songmass} applied the MASS~\cite{song2019mass} pre-training method to train a Transformer~\cite{waswani2017attention} model for the encoding and generation of symbolic melodies, which can be applied to downstream generation tasks. Wu et al.~\cite{wu2023melodyglm} introduced the widely used n-gram method from the NLP field into the melody generation domain, designing an n-gram pre-training method for symbolic melody generation models.

\subsection{Text}


Natural language text is the primary way humans describe the world. Similarly, a piece of music can also be richly described through text. Advances in NLP technology, particularly the development of LLMs in recent years, have provided abundant solutions and technical foundations for text-to-music generation.

In the field of NLP, numerous natural language encoding methods already exist. A classic approach involves using word embeddings, where models create a dictionary for each word, subword, or phrase and assign them embedding vectors in a latent space~\cite{mikolov2013vectorSpace,mikolov2013vector,pennington2014vector}. In deep NLP methods, encoders are often used to map raw data into a latent space. Consequently, pre-trained LM encoders can serve as language understanding models to encode natural language text into a latent space. Music generation models can then be trained based on this latent space to achieve text-guided music generation. BERT~\cite{devlin-etal-2019-bert} is a natural language encoding model built on this principle, constructed using a Transformer encoder. It obtains natural language encodings through pre-training and is applied to downstream tasks. T5~\cite{raffel2020t5} is a Transformer-based pre-trained text encoding model pre-trained on large-scale corpora, enabling transfer learning by reformulating downstream text processing tasks into a text-to-text format. FLAN-T5~\cite{chung2024flant5} further fine-tuned T5 on instruction datasets across multiple tasks, enhancing its generalization ability for zero-shot and few-shot tasks and improving its capacity to understand instructions. ERNIE-M~\cite{ouyang2021erniem} achieved cross-lingual semantic alignment to train multilingual encoding models using monolingual corpora.

Some works also explored joint audio-text embedding, mapping both modalities into the same latent space. These methods are usually based on contrastive learning, ensuring that semantically similar audio and text pairs are closer in the space, while dissimilar pairs are farther apart. Inspired by image-text embedding models like CLIP~\cite{radford2021clip}, Wav2CLIP~\cite{wu2022wav2clip} and AudioCLIP~\cite{guzhov2022audioclip} leveraged audio-text labeled data 
to build embedding models. CLAP~\cite{clap} extended this further by expanding text labels into natural language, creating a more flexible and scalable audio-text embedding model.

However, semantic gaps exist between professional music descriptions and colloquial language~\cite{wang2024muchin}. Directly applying LMs trained on general corpora to text-guided music generation may not suffice, highlighting the need for specialized music-centric natural language understanding models. MuLan~\cite{mulan} is a joint audio-text embedding model tailored for music, comprising an audio encoder and a natural language encoder, trained on dedicated music datasets. It maps music audio and text into the same latent space, where semantically related audio-text pairs are positioned closely, while dissimilar pairs are farther apart.

\subsection{Image}
\label{sec:imagerep}

Humans can relatively easily connect the sensations between vision and music, but it is not easy for AI to capture this connection.
Images are composed of pixels with continuous values. In contrast, musical notes are discrete, adhering to structured tuning systems, with each note carrying explicit attributes. This fundamental disparity in modality representation complicates direct cross-modal transfer between these two modalities. 

In the field of computer vision (CV), extensive research has produced diverse image encoding techniques. 
Since the advent of AlexNet~\cite{krizhevsky2012alexnet}, pretrained deep learning networks like ResNet~\cite{he2016resnet} have dominated feature extraction. 
Vision Transformer (ViT)~\cite{dosovitskiy2021vit}, on the other hand, has successfully applied Transformer from NLP to the field of CV and achieved remarkable results. This not only demonstrates the performance of Transformer~\cite{waswani2017attention} on images but also bridges the two modalities. As a result, multi-modal models can utilize ViT for image encoding.
In recent years, Latent Diffusion Models (LDMs)~\cite{rombach2022ldm} have achieved remarkable results in the field of image synthesis. MelFusion~\cite{melfusion} leverages LDM’s latent space to encode visual information, guiding diffusion models to generate music aligned with input images. This demonstrates the dual capability of LDMs as both image encoders and cross-modal facilitators, paving the way for unified multi-modal representations.

\subsection{Video}

Audio and video data dominate internet content, with videos often accompanied by music, making video-to-music generation a critical focus in multi-modal music generation. While video shares visual characteristics with static images, it inherently involves temporal continuity, comprising sequential frames that introduce a time dimension. Beyond extracting spatial features from individual frames, understanding video requires modeling temporal dynamics.  

The video consists of time-sequential frames, including both spatial and temporal information. Spatial features can be extracted using methods described in Section~\ref{sec:imagerep}. Temporal information is typically captured by sampling keyframes, keypoints, and optical flow. For instance, Simonyan and Zisserman~\cite{simonyan2014two} proposed a two-stream CNN architecture. Within the architecture, one stream processes static frames for spatial context, while the other analyzes optical flow across consecutive frames to model motion. 
Inspired by this, Feichtenhofer et al.~\cite{spatiotemporalresnet} adopted the two-stream architecture to build the Spatiotemporal Residual Networks. The model extracts appearance based on consecutive RGB images and motion based on optical flow. Furthermore, on this basis, it extends ResNet~\cite{he2016resnet} to the spatiotemporal domain, constructing 3D ResNet. Temporal information is extracted through 3D convolutions and residual connections. SlowFast Networks~\cite{feichtenhofer2019slowfast} represented a further development, in which a low frame rate 3D ResNet extracts spatial semantics and a high frame rate 3D ResNet extracts action information. 

Beyond CNNs, temporal modeling can also leverages sequential architectures like Transformer models. Arnab et al.~\cite{arnab2021vivit} proposed ViViT, adapting ViT~\cite{dosovitskiy2021vit} for video by splitting spatial and temporal encoding. Within the architecture, a spatial transformer extracts frame-level features, followed by a temporal transformer to model inter-frame relationships.  

To generate music aligned with video, the generated music must correspond to the rhythm, emotion, and semantics of the video. In the video, dynamic elements like motion and scene transitions often guide rhythmic patterns, while static features inform musical style, mood, and atmosphere. These extracted spatiotemporal representations serve as cross-modal anchors, enabling models to translate visual features into coherent auditory experiences.





\section{Modality-Oriented Music Generation}

Existing music generation techniques can derive guidance from multiple modalities, including auditory, textual, and visual inputs, to create corresponding music. It is evident that the content and representation methods of these guidance modalities vary significantly. Therefore, specialized techniques are required for each modality to maximize the utilization of its information and generate high-quality music that meets the desired criteria. This section will categorize and elaborate on the technical details of music generation methods tailored to different guidance modalities.

\subsection{Single-Modal Music Generation}
\label{sec:singlemodal}

Single-modal music generation leverages existing musical content to produce new compositions within the same modality, encompassing both symbolic-to-symbolic and audio-to-audio generation approaches. This section discusses part of the representative methods in this field.

\subsubsection{Audio-to-Audio Generation}
\label{sec:audio2audio}

Audio-based music generation typically requires models to generate music based on audio input, encompassing tasks such as synthesizing new sounds in the original audio domain, music continuation, inpainting, music editing, accompaniment generation, and more. 

VampNet~\cite{garcia2023vampnet} is a music audio encoder-decoder model built upon RVQ-VAE~\cite{zeghidour2021soundstream} and Transformer~\cite{waswani2017attention,shaw2018relative} architectures with flexible masking strategies. Utilizing the Improved RVQGAN~\cite{kumar2024improvedrvq} to encode audio into tokens, the researchers introduced Masked Visual Token Modeling~\cite{chang2022maskgit} into the audio domain to develop Masked Acoustic Token Modeling. This approach trains a non-autoregressive music generation model with strong flexibility and scalability. 
It leverages diverse masking patterns without conditional controls, enabling multiple tasks including music continuation, completion, inpainting, and periodic generation.
Based on the observation that music comprises contextually interrelated segments, Parker et al.~\cite{parker2024stemgen} developed StemGen, advancing the architectures of VampNet~\cite{garcia2023vampnet} and SoundStorm~\cite{borsos2023soundstorm}. Based on Masked Acoustic Token Modeling, they presented a framework featuring multi-channel audio embeddings and a novel sampling strategy for context-aware music generation. However, VampNet can only generate music clips up to 2 seconds. To address this, Lin et al.~\cite{airgen} developed AIRGen for flexible, controllable long-term music inpainting. By introducing heterogeneous adapters and masked training methods to finetune MusicGen~\cite{copet2024musicgen}, AIRGen achieves a masked LM capable of longer and more controllable music completion.

While the Music VQ-VAE~\cite{dhariwal2020jukebox,jukemir} mentioned in Section~\ref{sec:audio} demonstrates good audio modeling performance, its compression and reconstruction still incurs audio information loss, which leads to artifacts. To comprehensively model audio information, Borsos et al.~\cite{borsos2023audiolm} developed AudioLM, concurrently employing SoundStream~\cite{zeghidour2021soundstream} and w2v-BERT~\cite{w2vbert} to extract acoustic and semantic tokens respectively. The researchers found these two types of tokens complementary in audio discrimination and reconstruction quality. Combined, they enable high-quality piano music continuation. 
However, given its primary nature as a speech synthesis model, AudioLM's musical applications remain underexplored. Donahue et al.~\cite{donahue2023singsong} adapted AudioLM for conditional music generation, specifically vocal-to-accompaniment generation. 

\subsubsection{Symbolic-to-Symbolic Generation}

Symbolic music generation typically takes symbolic music as input and requires models to generate corresponding music based on that input. Similar to audio-based approaches, common tasks include inpainting, continuation, composition, adaptation and accompaniment generation.

A representative work 
is MusicVAE~\cite{roberts2018musicvae}. While traditional VAEs~\cite{kingma2014vae1,rezende2014vae2} excel at encoding continuous data with fixed dimensionality such as images, they struggle with discrete sequential data such as symbolic music. To adapt VAEs for symbolic music encoding and generation, the researchers introduced a hierarchical decoder into the recurrent VAE architecture. 
This hierarchical approach decodes long sequences through subsequences, effectively addressing the challenges of long sequence decoding. MusicVAE enables encoding and reconstruction of long-term symbolic music sequences, as well as generating new symbolic music by interpolating different latent space encodings.

Music continuation and inpainting require models to generate new content based on existing musical material to fill the blanks in the music, demanding strong structural awareness. Models must develop musical ideas through repetition and variation rather than simple replication. To address limitations of foundational autoregressive models in learning musical structure, Shih et al.~\cite{shih2022theme} introduced theme-based conditioning and proposed Theme Transformer. This approach uses contrastive learning to identify musical themes and explicitly conditions the Transformer to repeatedly incorporate these themes in generated outputs. Wu et al.~\cite{wu2023melodyglm} introduced a multi-task pre-training framework based on the n-gram method, achieving the capability to generate melodies with long-term structure through large-scale pretraining. Mittal et al.~\cite{mittal2021symbolic} applied diffusion models to symbolic music generation by using MusicVAE~\cite{roberts2018musicvae} as a latent encoder. By conditioning on surrounding musical segments and incorporating Transformer-based inter-bar coherence control, the model effectively generates missing middle sections of music.

While music continuation and inpainting focus on temporal generation, accompaniment generation requires harmonic generation, modeling relationships between simultaneous notes. 
Zhu et al.~\cite{zhu2018xiaoice} presented an end-to-end framework for melody and accompaniment generation. Through multitask learning, the system generates melodies and accompaniment while enabling multi-instrument harmonies. Building on this, the researchers integrated style discriminators 
to enable genre-specific generation~\cite{zhu2020xiaoice2}. For long-term accompaniment generation, Ren et al.~\cite{ren2020popmag} introduced PopMAG, along with MuMIDI, an innovative multi-track MIDI representation. This representation encodes multiple instruments into a unified sequence while explicitly capturing inter-track dependencies. To address sequence length challenges from MuMIDI, the researchers compressed the representations and introduced long-term memory mechanisms inspired by Transformer-XL~\cite{dai2019transformerxl}. For real-time applications, Wang et al.~\cite{wang2022songdriver} implemented a two-stage generation system. This method distills Transformer~\cite{waswani2017attention} with a CRF~\cite{lafferty2001crf} model to maintain high generation quality while keeping low physical latency. The researchers later extended the work towards multi-modal music generation, incorporating emotion control through downsampling and semi-supervised learning for smooth affective music adaptation~\cite{wang2024remast}.

\subsection{Cross-Modal Music Generation}

\begin{table*}[t]
\centering
\caption{Cross-Modal Music Generation Methods}
\label{tab:cross-modal}
\resizebox{\textwidth}{!}{
\begin{tabular}{lllllll}
\toprule
\multirow{2}{*}{\textbf{Task}} & \multirow{2}{*}{\textbf{Method}} & \multirow{2}{*}{\textbf{Year}} & \multicolumn{4}{c}{\textbf{Modeling Approach}} \\
\cmidrule(lr){4-7}
& & & \textbf{Input Encoding} & \textbf{Output Encoding} & \textbf{Cross-Modal Method} & \textbf{Model Architecture} \\
\midrule
\multirow{8}{*}{Score-to-Audio} & Raphael~\cite{raphael2009representation} & 2009 & Symbolic annotations & Audio parameters & Rule-based mapping & Markov model \\
& Manzelli et al.~\cite{manzelli2018conditioning} & 2018 & MIDI + LSTM & Waveform & Mapping & WaveNet~\cite{van2016wavenet} \\
& Mel2Mel~\cite{kim2019mel2mel} & 2019 & Onsets and Frames~\cite{hawthorne2018onsets} & Mel spectrogram & Mapping & Bi-LSTM + WaveNet~\cite{van2016wavenet} \\
& PerformanceNet~\cite{wang2019performancenet} & 2019 & Piano roll + U-Net & STFT & Mapping & PGGAN~\cite{karras2018pggan} \\
& MIDI-DDSP~\cite{wumididdsp} & 2022 & MIDI & Mel spectrogram & Mapping & RNN + DDSP~\cite{engelddsp} \\
& Deep Performer~\cite{dong2022deepPerformer} & 2022 & Note attributes & Mel spectrogram & Embedding & Transformer \\
& Hawthorne et al.~\cite{hawthorne2022multi} & 2022 & MIDI & Mel spectrogram & Cross-attention & T5-based DDPM \\
& Maman et al.~\cite{maman2024performance} & 2024 & MIDI & Mel spectrogram & Mapping & FiLM+ DDPM \\
\midrule
\multirow{3}{*}{Text-to-Symbolic} & BUTTER~\cite{zhang2020butter} & 2020 & Keywords & One-hot vectors & Joint embedding & GRU \\
& Wu and Sun~\cite{wu2023exploring} & 2023 & Pre-trained LMs & ABC & Sequence mapping & Pre-trained LMs \\
& MuseCoCo~\cite{lu2023musecoco} & 2023 & BERT & REMI & Attribute bridging & Linear Transformer~\cite{katharopoulos2020transformers} \\
\midrule
\multirow{5}{*}{Text-to-Audio} & MusicLM~\cite{agostinelli2023musiclm} & 2023 & MuLan & SoundStream + w2v-BERT & Joint embedding & Transformer \\
& Noise2Music~\cite{huang2023noise2music} & 2023 & T5 & Waveform / Log-Mel & Cross-attention & LDM \\
& Moûsai~\cite{mousai-2024-efficient} & 2024 & T5 & Diffusion-based autoencoder & Cross-attention & LDM \\
& MusTango~\cite{melechovsky2024mustango} & 2024 & FLAN-T5 + Embedding & VAE & Cross-attention & LDM \\
& MeLoDy~\cite{lam2024MeLoDy} & 2024 & MuLan & Wav2Vec2-Conformer~\cite{baevski2020wav2vec,gulati2020conformer} + VAE-GAN & Joint embedding & LLaMA + LDM \\
& AudioLDM 2~\cite{liu2024audioldm2} & 2024 & CLAP + FLAN-T5 & AudioMAE & Cross-attention & GPT-2 + LDM \\
& MuDiT \& MuSiT~\cite{wang2024mudit} & 2024 & Joint embedding model & Mel spectrogram & Joint embedding & DiT / SiT \\
\midrule
\multirow{7}{*}{Lyrics-to-Melody} & T-Music~\cite{long2013tmusic} & 2013 & Tone & Note pitch & Rule-based mapping & Probabilistic automaton \\
& iComposer~\cite{lee2019icomposer} & 2019 & LSTM & Note pitch and duration & Mapping & LSTM \\
& Yu et al.~\cite{yu2021lstmgan} & 2021 & Skip-gram models & Note pitch and duration & Mapping & LSTM with adversarial learning \\
& SongMASS~\cite{sheng2021songmass} & 2021 & Token sequence & Note pitch and duration & Cross-attention & Transformer \\
& TeleMelody~\cite{ju2022telemelody} & 2022 & Template & Template & Template bridging & Transformer \\
& ReLyMe~\cite{zhang2022relyme} & 2022 & Tone, rhythm and structure & Tone, rhythm and structure & Rule-based mapping & SongMASS / TeleMelody \\
& SongGLM~\cite{yu2024songglm} & 2025 & Syllable stress & Event token sequence & Concatenation & GLM \\
\midrule
\multirow{2}{*}{Image-to-Symbolic} & SDMuse~\cite{sdmuse} & 2023 & SDE & Piano roll + MIDI & Joint embedding & DDPM + CNN + Transformer-XL \\
& Drawlody~\cite{drawlody} & 2024 & CNN & Proposed FlexMIDI & Cross-attention & Transformer-XL \\
\midrule
\multirow{1}{*}{Image-to-Audio} & Wang et al.~\cite{wang2023continuous} & 2023 & ALAE~\cite{pidhorskyi2020alae} / VQ-VAE / $\beta$-VAE~\cite{higgins2017beta} & FNT~\cite{tan2020fnt} / LSR~\cite{pati2019lsr} & Emotion bridging & FNT~\cite{tan2020fnt} / LSR~\cite{pati2019lsr} \\
\midrule
\multirow{7}{*}{Video-to-Symbolic} & Foley Music~\cite{gan2020foleymusic} & 2020 & OpenPose + ST-GCN & MIDI & Cross-attention & Transformer decoder \\
& RhythmicNet~\cite{su2021rhythmic} & 2021 & \makecell[l]{OpenPose + ST-GCN + \\ Transformer encoder} & REMI & Mapping & Transformer + U-Net \\
& CMT~\cite{di2021cmt} & 2021 & Optical flow + Visual beats & CP & Rule-based mapping & Linear Transformer~\cite{katharopoulos2020transformers} \\
& V-MusProd~\cite{zhuo2023video} & 2023 & \makecell[l]{CLIP2Video~\cite{fang2021clip2video} + \\ 2D feature map~\cite{afifi2021histogan} + \\ RGB difference} & Event token sequence & Cross-attention & Transformer \\
\midrule
\multirow{4}{*}{Video-to-Audio} & MI-Net~\cite{su2020minet} & 2020 & OpenPose + Bi-GRU & VQ-VAE & Cross-attention & Transformer decoder \\
& D2M-GAN~\cite{zhu2022d2mgan} & 2022 & SMPL + 2D body keypoints+ I3D & Music VQ-VAE & Mapping & GAN \\
& LORIS~\cite{yu2023loris} & 2023 & I3D + Bi-LSTM + Poses & LDM-based encoder & Cross-attention & LDM \\
& Li et al.~\cite{li2024dance} & 2024 & Keypoints & VQ-VAE / VAE & Text bridging & Text-to-music models \\
\bottomrule
\end{tabular}
}
\end{table*}

Cross-modal music generation exploits cross-modal relationships to produce conditionally consistent music. This section categorizes approaches by input modalities, focusing on score-to-audio, text-to-music, and visual-to-music tasks, while analyzing cross-modal mapping strategies. Some representative methods are shown in Table~\ref{tab:cross-modal}.

\subsubsection{Score-to-Audio Generation}

Symbolic music cannot be directly perceived as audible sound by humans. Consequently, generating audio from musical scores has emerged as a key research direction in music generation. 

Raphael~\cite{raphael2009representation} developed an expressive melody synthesis system by simulating human performance. 
This system utilizes a Markov model to generate symbolic annotations for each note, and maps these annotations to audio parameters.

With advances in deep learning, Manzelli et al.~\cite{manzelli2018conditioning} adapted WaveNet~\cite{van2016wavenet} for score-to-audio synthesis, while Kim et al.~\cite{kim2019mel2mel} enhanced WaveNet with timbre embeddings and FiLM layers for flexible timbre control. However, as a general audio synthesis model, WaveNet lacks specialized design for polyphonic structures of music, complicating the synthesis of harmonically rich audio. 
To address this, Wang and Yang~\cite{wang2019performancenet} introduced a two-stage framework. In the first stage, a U-Net~\cite{ronneberger2015unet} architecture learns the mappings between piano roll representations and spectral features. Based on this, a texture refinement network is employed to enhance harmonic details. 



Ensuring expressiveness and realism in synthesized audio, Wu et al.~\cite{wumididdsp} further enhanced controllability, allowing users to comprehensively control details at various levels. However, these previous works overlooked the impact of expressive timing in real performances on musical expressiveness. To address this issue, Dong et al.~\cite{dong2022deepPerformer} introduced a Transformer-based alignment model to predict performance-specific onsets and durations in actual musical performances, thereby synthesizing expressive musical performance audio. 

While existing systems typically focus on single-instrument generation, Hawthorne et al.~\cite{hawthorne2022multi} developed a T5-based~\cite{raffel2020t5} diffusion model~\cite{ho2020ddpm} trained on multi-instrument datasets to generate multi-track audio from MIDI scores. To enhance realism, Maman et al.~\cite{maman2024performance} later incorporated performance conditioning via FiLM layers~\cite{perez2018film} into diffusion models, enabling realistic audio synthesis with improved control of timbre and style.

\subsubsection{Text-to-Music Generation}
\label{sec:text2music}

Text-to-music generation requires models to generate music based on natural language text. 
The text can be a general description, as well as musical terms. Additionally, the text can also be lyrics, where the model must generate music that fits the style and prosody of the lyrics. Currently, text-to-music generation mainly involves two technical approaches: symbolic music generation and audio-based music generation. Furthermore, generating melodies from lyrics is also a key task within text-to-music generation.

In the field of \textbf{symbolic music generation}, 
Zhang et al.~\cite{zhang2020butter} implemented BUTTER, decomposing symbolic music into attribute subspaces aligned with text keywords in a VAE latent space. While it successfully maps specific keywords to corresponding musical attributes, the framework's dependence on predefined keywords limits its ability to handle natural language descriptions in practical applications.
Wu and Sun~\cite{wu2023exploring} regarded text-to-music generation as a sequence-to-sequence task, leveraging pre-trained LMs for text-to-music generation. Compared to BUTTER, this approach supports more flexible text inputs. However, it suffers from limited musical quality and diversity due to small-scale training data.
Lu et al.~\cite{lu2023musecoco} addressed these limitations through a two-stage framework involving text-to-attribute transformation and subsequent attribute-to-music generation. The researchers constructed a large-scale dataset based on this two-stage design. MIR techniques were employed to establish attribute-music correlations. And ChatGPT~\footnote{\url{https://openai.com/chatgpt/overview}} was leveraged to create text-to-attribute mapping frameworks. This two-stage design enables training large models on extensive data, improving generation quality.

In the field of \textbf{audio-based music generation}, a large number of text-to-music generation algorithms have been proposed in recent years. 
Developed concurrently, Moûsai~\cite{mousai-2024-efficient} and Noise2Music~\cite{huang2023noise2music} both leverage LDMs~\cite{rombach2022ldm} for text-conditioned music generation. Both models employ T5~\cite{raffel2020t5} for text understanding. Moûsai utilizes waveforms as audio representations and encodes audio based on a diffusion autoencoder~\cite{preechakul2022diffusion}. 
In contrast, Noise2Music is a two-stage framework, and it explores both waveform and log-Mel spectrogram representations. Noise2Music highlights the efficacy of cascaded models in extracting semantic information from text for high-fidelity audio synthesis.
Melechovsky et al.~\cite{melechovsky2024mustango} used FLAN-T5~\cite{chung2024flant5} as the text understanding model and trained a Music-Domain-Knowledge-Informed U-Net (MuNet) as the denoising model for the LDM, achieving audio-based music generation from text instructions. Currently, Diffusion Transformer (DiT)~\cite{peebles2023dit} and its further development, Scalable Interpolant Transformer (SiT)~\cite{ma2024sit}, have achieved good results in image and video generation. They can also be applied to music generation. Wang et al.~\cite{wang2024mudit} developed MuDiT \& MuSiT based on DiT and SiT. Inspired by CLAP~\cite{clap} and MuLan~\cite{mulan}, the researchers trained a music-text jointly embedding model to encode the colloquial music description. 
Finally, they applied DiT and SiT to construct two music audio generation models, achieving music generation from colloquial descriptions.

Besides diffusion models, Transformer-based LMs can also be applied to audio-based music generation. AudioLM~\cite{borsos2023audiolm} successfully introduced LMs to model audio generation tasks, which is introduced in Section~\ref{sec:audio2audio}. Recognizing the potential of AudioLM for music generation, Agostinelli et al.~\cite{agostinelli2023musiclm} expanded it by introducing text as a control condition, leading to MusicLM. MusicLM is a cascaded model with three stages. First, SoundStream~\cite{zeghidour2021soundstream} encodes audio through quantization. Second, w2v-BERT~\cite{w2vbert} extracts semantic information from the audio itself. Third, MuLan~\cite{mulan} ensures semantic alignment between input text and the generated music audio.


Combining diffusion and LMs is also a typical architecture in music generation. In this approach, the LM is employed to model the semantic features of audio, thereby providing guidance over audio generation at the semantic level. Subsequently, diffusion models, with their advantage in generating audio details, serve as the audio synthesis module. This module effectively integrates semantic information into the final audio output.
To address the limitations of MusicLM~\cite{agostinelli2023musiclm} in terms of music generation efficiency, Lam et al.~\cite{lam2024MeLoDy} proposed MeLoDy.
Building upon the approach of utilizing a high-level LM~\cite{touvron2023llama} for audio semantic modeling, the researchers further developed the architecture by incorporating a dual-path diffusion model and an audio VAE-GAN~\cite{kingma2014vae1,rezende2014vae2,kong2020hifi,jang2021univnet}. These two models decode semantic tokens into audio, improving generation efficiency and quality.
To establish a unified audio generation framework, Liu et al. proposed AudioLDM 2~\cite{liu2024audioldm2}. In this framework, AudioMAE~\cite{huang2022audiomae} is utilized to extract audio semantic features, thereby establishing the Language of Audio (LOA) as a universal audio representation.  A fine-tuned GPT-2 model~\cite{radford2019gpt2} is employed to generate LOA from textual input. Subsequently, a self-supervised LDM~\cite{rombach2022ldm} generates audio under the conditional control of LOA, based on the audio latent space of a VAE~\cite{kingma2014vae1}.

\textbf{Lyrics-to-melody generation} is also one of the tasks in text-to-music generation in addition to generating music based on text descriptions. As early as 2013, Long et al.~\cite{long2013tmusic} recognized the importance of leveraging the correlation between lyrics and melodies in music generation tasks and proposed T-Music. They mined frequent patterns between lyrics and melodies in a song database and constructed an automatic probabilistic automaton based on the data mining results. However, this early research struggles with issues such as poor performance and monotonous melody generation. With the maturation of deep learning, many algorithms for lyrics-to-melody generation based on deep learning methods have been proposed. Lee et al.~\cite{lee2019icomposer} proposed iComposer, an LSTM-based~\cite{hochreiter1997lstm} songwriting system for Chinese popular music. iComposer consists of two independent LSTM models that generate rhythms and pitches from lyrics respectively. However, the small number of lyrics-melody data pairs limits the scale and performance of deep generative models for lyrics-to-melody generation. Yu et al.~\cite{yu2021lstmgan} constructed a large-scale dataset, including 12,197 songs in MIDI format. In this research, the researchers introduced the conditional Generative Adversarial Network (GAN)~\cite{goodfellow2014gan,mirza2014cgan} to the LSTM model, constructing a conditional LSTM-GAN model to achieve the generation of melodies from lyrics. In terms of text encoding, two independent skip-gram models are trained to obtain text embedding vectors at the word level and syllable level respectively. And then the two levels of vectors are concatenated as the overall text embedding vector. Despite researchers continuously expanding the scale of data, constructing a dataset that aligns lyrics and melodies is still not an easy task. Deep generative models in this field are still limited by the lack of corresponding data. To address this issue, Sheng et al.~\cite{sheng2021songmass} constructed a pre-training method SongMASS based on the sequence-to-sequence pre-training method MASS~\cite{song2019mass}. Through this method, the lyrics encoder and melody encoder can be pre-trained using unpaired lyrics and melodies, thus reducing the demand for lyrics-melody data pairs. But SongMASS still needs to use some lyrics-melody data pairs to learn the association between lyrics and melodies. Unlike the end-to-end architecture in SongMASS, Ju et al.~\cite{ju2022telemelody} proposed a two-stage model, TeleMelody, to eliminate the dependence on lyrics-melody data pairs. In TeleMelody, there is an intermediate state called templates between lyrics and melodies. The template includes musical elements such as tonality, rhythm patterns, chords, and rhyming. These musical elements can be extracted from music using MIR methods to construct template-melody pairing data. Lyrics-template pairs can be extracted from the widely available karaoke data on the internet. Based on these two datasets, two models for lyrics-template and template-melody can be trained separately. And then the two models can be concatenated to achieve lyrics-melody generation. However, the relationship between lyrics and melodies is complex and subtle. Despite learning from a large amount of data, models still struggle to capture this complexity, leading to conflicts between generated melodies and lyrics. To address this issue, Zhang et al.~\cite{zhang2022relyme} proposed ReLyMe, considering lyrics-melody relationships from the perspectives of linguistics and music theory. This method adds constraints during the decoding stage, achieving harmony between generated melodies and lyrics.
However, previous works usually suffer from two main challenges, including one-to-many syllable-note alignment and lyrics-melody harmony modeling. Intermediates like Telemelody~\cite{ju2022telemelody} or strict rules like ReLyMe~\cite{zhang2022relyme} limit models' capabilities and generative diversity. To address these key limitations in existing approaches, Yu et al.~\cite{yu2024songglm} introduced SongGLM, a novel lyrics-to-melody generation system. 
In SongGLM, lyrics tokens and musical event tokens are concatenated. For lyrics-melody alignment, SongGLM directly captures hierarchical relationships between lyrics and corresponding melodic phrases through a 2D alignment encoding scheme. For lyrics-melody harmony, a harmonized n-gram extraction method is presented to leverage relationships between lyrics features and melodic features. Furthermore, the researchers introduced a multi-task pre-training framework based on the General Language Model (GLM)~\cite{du2022glm} to train the model at various levels.

\subsubsection{Visual-to-Music Generation}

Visual-to-music generation requires models to produce corresponding music based on visual information, primarily conveyed through images and videos. 
Currently in its nascent stage, research on visual-to-music generation remains insufficient and requires further exploration and development.

\textbf{Image-to-music generation} requires models to create music that matches the atmosphere of scenes depicted in images. In recent studies on music generation, researchers have explored using simple images as conditional controls for model outputs. For instance, SDMuse~\cite{sdmuse} proposed by Zhang et al. generates symbolic music aligned with stroke patterns from input sketch images, demonstrating that visual information can guide music generation to some extent. SDMuse encodes both images and music into Gaussian noise within a shared latent space through a stochastic differential equation (SDE)~\cite{song2021sde}. By diffusing image inputs into noise and reconstructing them through denoising, the model generates music that adheres to visual patterns. Similiarly, Drawlody~\cite{drawlody} generates melodies guided by sketches. It introduces a generalized representation named FlexMIDI for sketch-based melody generation, and employs a CNN for sketch encoding. However, sketches are simplistic visual carriers with limited information, differing significantly from realistic artworks. 
In response to the rich emotional expressions in images, Wang et al.~\cite{wang2023continuous} proposed a continuous emotion-based image-to-music generation framework, which uses emotion as a bridge for cross-modal generation.
Approaches like MelFusion~\cite{melfusion} and MuMu-LLaMA~\cite{liu2024m2ugen} employ distinct multi-modal strategies to achieve music generation guided by images and other modalities, which will be detailed in Section~\ref{sec:multi-modal}.

\textbf{Video-to-music generation} requires models to generate appropriate music with a given video, demanding both content consistency and temporal alignment. 
For musical performance videos, Gan et al.~\cite{gan2020foleymusic} developed a Graph-Transformer network, Foley Music. Foley Music encodes videos via OpenPose keypoints~\cite{cao2020openpose} and a spatio-temporal graph convolutional network (ST-GCN)~\cite{yan2018stgcn}, and then generates MIDI-format music via a Transformer decoder~\cite{huang2018musictrm,waswani2017attention}. However, Foley Music includes limitations of training a distinct model for each individual instrument and requires instrument labels.
Su et al. proposed MI-Net, an unsupervised model using bidirectional GRUs and VQ-VAE to generate music from performance videos of different instruments without labels. 

Recently, some studies have explored the generation of music based on dance videos. Dance music generation requires multi-dimensional alignment between music and dance in rhythm, style, emotion, and other aspects. 
For rhythmic alignment, Su et al.~\cite{su2021rhythmic} proposed RhythmicNet, a three-stage model for symbolic music generation based on motion rhythms. The model sequentially processes motion-to-music generation through Video2Rhythm, Rhythm2Drum, and Drum2Music stages. It extracts motion key points~\cite{cao2020openpose} to predict beat frames, generates corresponding drum patterns, and finally incorporates additional instrument tracks to produce complete music synchronized with the original human movements.
However, this symbolic music generation approach exhibits limitations in terms of instrumental richness and may not adequately accommodate the stylistic diversity inherent in dance videos. 
Zhu et al.~\cite{zhu2022d2mgan} proposed an audio-based dance-to-music generation model, D2M-GAN, to overcome these limitations.
For video encoding, D2M-GAN employs the 3D Skinned Multi-Person Linear model (SMPL)~\cite{loper2015smpl} in conjunction with 2D body keypoints~\cite{cao2020openpose} to extract human motion features and utilizes I3D features~\cite{carreira2017i3d} to encode visual information from the video. For audio generation, D2M-GAN adopts Music VQ-VAE~\cite{dhariwal2020jukebox} as the audio codec model and constructs a GAN-based model for audio token generation. Ultimately, D2M-GAN demonstrates the capacity to generate complex and richly textured musical audio that is responsive to the rhythm and content of dance videos.
Nonetheless, training such models from scratch necessitates a large scale of paired video-music data. Given the scarcity of high-quality paired datasets, the exploration of pre-trained music generation models to alleviate the data demands of model training presents a salient research direction.  
Li et al.~\cite{li2024dance} developed an encoder-based textual inversion technique that integrates rhythm and style information from dance videos into pre-trained text-to-music models. Their Dual-Path Rhythm-Genre Inversion method extracts rhythmic patterns and stylistic elements via specialized encoders, mapping these features to text embeddings while dynamically adjusting pseudo-word representations to enable precise alignment with diverse dance movements.
By incorporating a pre-trained music generation model, the data volume required for training is reduced to approximately 1\% of that needed for training from scratch. 

However, generating music for dance videos is still somewhat distinct from generating music for general videos. 
Yu et al.~\cite{yu2023loris} proposed the LORIS framework, broadening the applicability of models from dance to diverse sports scenarios, such as gymnastics and figure skating. LORIS is a generative framework based on a latent conditional diffusion model~\cite{rombach2022ldm,karras2022edm}, achieving long-term music generation synchronized with the video rhythm. Specifically, LORIS initially extracts visual and rhythmic information from videos through a context-aware conditional encoder. Within the encoder, visual information is captured for temporal dependencies using a pre-trained I3D network~\cite{carreira2017i3d} and bidirectional LSTM~\cite{hochreiter1997lstm}, while rhythmic information is extracted from motions through an improved rule-based method. Subsequently, these multi-modal information streams are integrated into a latent diffusion model via a cross-modal attention mechanism. The framework further employs the Hawkes process~\cite{hawkes1971spectra,mei2017neuralhawkes,zhang2020selfhawkes} to enhance positional encoding of rhythmic points, ensuring a high degree of consistency between the generated music and the video rhythm. Nevertheless, this work remains confined to videos of human motions.
Di et al.~\cite{di2021cmt} introduced the Controllable Music Transformer (CMT), which generates symbolic music that aligns with video rhythm. CMT establishes correlations between video elements and musical attributes. By incorporating these video-derived attributes along with user-specified instrument and style preferences, it produces customized background music for videos.
However, CMT extracts and utilizes only rhythmic information from videos, representing a limited fraction of the information contained within videos. 
To advance research in video background music generation, Zhuo et al.~\cite{zhuo2023video} proposed a comprehensive solution encompassing a dataset, benchmark model, and evaluation metrics. The researchers presented a Transformer-based three-stage generative model, V-MusProd, sequentially generating chords, melodies, and accompaniment.  Beyond referencing the methods in CMT to control music rhythm using motion features, V-MusProd also extracts semantic and color information from videos to generate music consistent with the video themes and emotions. 
However, V-MusProd still exhibit limitations in fully utilizing video information. Video background music generation necessitates a unified model capable of comprehensively understanding complete videos and generating corresponding music. This calls for multi-modal generative models, which will be detailed in Section~\ref{sec:multi-modal}.

\subsection{Multi-Modal Music Generation}
\label{sec:multi-modal}

\begin{table}[t]
\centering
\caption{Multi-Modal Music Generation Methods}
\label{tab:multimodal_music}
\begin{threeparttable}
\resizebox{\textwidth}{!}{%
\begin{tabular}{llllll}
\toprule
\multirow{2}{*}{\textbf{Method}} & \multirow{2}{*}{\textbf{Year}} & \multirow{2}{*}{\textbf{Guidance Modalities}} & \multicolumn{3}{c}{\textbf{Modeling Approach}} \\
\cmidrule(lr){4-6}
& & & \textbf{Modal Encoding} & \textbf{Modal Fusion} & \textbf{Model Architecture} \\
\midrule
\multicolumn{6}{c}{\textit{Single External Modality}} \\
\midrule
Jukebox~\cite{dhariwal2020jukebox} & 2020 & Text + Audio → Audio & Transformer encoder + Music VQ-VAE & Cross-attention & Sparse Transformer~\cite{child2019sparsetrm} \\
MusicGen~\cite{copet2024musicgen} & 2024 & Text + Audio → Audio & T5 + EnCodec + Chromagram & Cross-attention & Transformer decoder \\
AP-Adapter~\cite{tsai2024apadapter} & 2024 & Text + Audio → Audio & FLAN-T5 + CLAP + AudioMAE & Cross-attention & AudioLDM 2 \\
\midrule
\multicolumn{6}{c}{\textit{Multiple External Modalities}} \\
\midrule
MelFusion~\cite{melfusion} & 2024 & Text + Image → Audio & T5 + LDM + VAE & Cross-attention & LDM \\
Diff-BGM~\cite{li2024diffbgm} & 2024 & Text + Video → Symbolic music & BERT + VideoCLIP~\cite{xu2021videoclip} + Piano roll & Cross-attention + Feature selection & LDM \\
Seed-Music~\cite{bai2024seedmusic} & 2024 & Text + Audio + Symbolic → Audio & MuLan + VQ-VAE + REMI + Lead sheet & Multi-stage conditioning\tnote{*} & LM + DiT \\
MuMu-LLaMA~\cite{liu2024m2ugen} & 2024 & Text + Image + Video → Audio & LLaMA + ViT + ViViT + MERT & LLM bridging & LLaMA + MusicGen \\
V2Meow~\cite{su2024v2meow} & 2024 & Text + Video → Audio & \makecell[l]{I3D + CLIP + ViT-VQGAN + \\ MuLan + w2v-BERT + SoundStream} & Concatenation & Transformer \\
XMusic~\cite{xmusic2025} & 2025 & \makecell[l]{Text + Image + Video + Audio \\ → Symbolic music} & \makecell[l]{Sentence-BERT~\cite{reimers2019sentence} + \\ ResNet + CLIP + \\ Rule-based methods + CP} & Concatenation & Transformer decoder \\
\bottomrule
\end{tabular}%
}
\begin{tablenotes}
\item[*] Including prefix conditioning, cross-attention conditioning, and temporal conditioning.
\end{tablenotes}
\end{threeparttable}
\end{table}

Multi-modal music generation leverages the integration of diverse modalities to guide the generative process. As discussed in Section~\ref{sec:intro}, these modalities may include both internal musical and external modalities. 
Notably, symbolic music and audio should be regarded as external modalities to each other due to their distinct representations and characteristics.
Multi-modal music generation can be categorized based on the number of external modalities included in the guidance modalities.
Some representative methods are shown in Table~\ref{tab:multimodal_music}.

\subsubsection{Multi-Modal Music Generation Guided by a Single External Modality}

Music generation with single external modal guidance utilizes internal modalities combined with a single external modality to jointly guide the generation process. The internal modalities provide musical prompts to the model, while the external modality offer additional conditions and controls to enhance controllability. This type of multi-modal music generation approaches include applications such as music adaptation, accompaniment-based singing synthesis, and music generation guided by musical prompts. 

\textbf{Jukebox}~\cite{dhariwal2020jukebox} is a model for generating diverse high-fidelity music in the raw audio domain, taking both audio and text as input modalities. In Jukebox, input audio is encoded into quantized representations via Music VQ-VAE, while text inputs including artist, genre and lyrics serve as conditional controls. A Transformer model processes these inputs, and the VQ-VAE decoder reconstructs the final audio. Benefiting from its high-fidelity encoding, conditional control mechanisms, and pretraining techniques, Jukebox achieves high-quality vocal synthesis in raw audio with controllable artistic styles and lyrics via text inputs.  

\textbf{AP-Adapter}~\cite{tsai2024apadapter} is a fine-tuning strategy for existing generative models, which adds audio as an input modality to text-to-music models. By leveraging textual control, it enables music adaptation based on input audio, achieving strong results in timbre conversion, style transfer, and accompaniment generation.  

\textbf{MusicGen}~\cite{copet2024musicgen} is a unified end-to-end model inspired by MusicLM, which streamlines the cascaded structure and employs an enhanced codebook strategy for more efficient generation. Beyond text conditioning, MusicGen can use melody as an additional control signal. It employs a chromagram-based melody conditioning method, leveraging information bottlenecks and unsupervised learning to enable music generation guided by both text and audio modalities.

However, such methods solely rely on a single external modality to guide the transformation of internal modalities, which is insufficient for exploring multi-modal interactions.

\subsubsection{Multi-Modal Music Generation Guided by Multiple External Modalities}

Music generation with multiple external modal guidance emphasizes the fusion of information from multiple external modalities and their alignment with music. The model must fully understand and integrate diverse external modalities to generate music that adheres to multi-modal semantics and control conditions.  

\textbf{Seed-Music}~\cite{bai2024seedmusic} is a unified multi-modal music generation system integrating text, audio, and symbolic modalities. For text encoding, the system encodes textual descriptions via MuLan and represents lyrics as phonemes. For audio encoding, it uses LM-based audio tokenizers. For symbolic encoding, it employs REMI~\cite{remi} as notation and uses lead sheets as tokens. Seed-Music comprises three complementary pipelines working together. The LM-based pipeline uses Transformer to generate audio tokens which are then rendered by a DiT model~\cite{peebles2023dit}. The symbolic pipeline generates and renders lead sheet tokens instead, enhancing interpretability and editing capabilities. The vocoder pipeline directly generates audio from input conditions using DiT. By combining these pipelines, Seed-Music integrates text, audio, and symbolic modalities while enabling user editing. The study also compares LM-based and diffusion-based pipelines, finding LM-based approaches more suitable for user interaction and future integration with other multi-modal models. Although Seed-Music integrates text, symbolic music, and audio modalities, it lacks guidance from the visual modality.

\textbf{MelFusion}~\cite{melfusion} is a diffusion-based music generation model prompted by both text and images. It extracts image semantic features using a pre-trained text-to-image LDM~\cite{rombach2022ldm}. These features are then infused into the text-to-music LDM via cross-attention between the decoder layers of both LDMs. The text-to-music LDM combines text inputs and image information to produce music aligned with both text and visual content. The experimental results demonstrate that integrating visual information into text-to-music generation improves the quality of generated music and enhances its alignment with human perception.  

\textbf{Diff-BGM}~\cite{li2024diffbgm} is a novel diffusion-based framework designed for video background music generation, which incorporates text and video modalities. Specifically, dynamic video features are utilized to influence the rhythmic elements of the music, while semantic features, extracted from video captions, guide the melody and overall atmospheric qualities. To achieve temporal coherence between the video and the generated music, a segment-aware cross-attention layer is incorporated, facilitating sequential alignment by focusing on short-term contextual relationships of modalities.

\textbf{V2Meow}~\cite{su2024v2meow} is a multi-modal music generation model that generates music from video and text inputs. It confronts the challenge of learning globally aligned signatures between video and music directly, without explicitly modeling domain-specific rhythmic or semantic relationships. For frame-level video understanding, it employs I3D~\cite{carreira2017i3d} and CLIP~\cite{radford2021clip} for visual feature extraction, and adopts ViT-VQGAN~\cite{yu2022vitvqgan} to obtain discrete video tokens. Similar to MusicLM~\cite{agostinelli2023musiclm}, V2Meow utilizes w2v-BERT~\cite{w2vbert} and SoundStream~\cite{zeghidour2021soundstream} for both semantic and acoustic encoding of audio. Additionally, it leverages MuLan~\cite{mulan} for audio-text embedding, and concatenates these embeddings with the video tokens for high-level control guided by text.

While these works jointly leverage visual and textual modalities to guide music generation, they lack a unified framework to integrate additional modalities. 
MelFusion solely employs images to provide visual information, without incorporating video. Although Diff-BGM integrates both video and textual information, the textual modality is merely used to express video semantics, resulting in insufficient exploration of text. Furthermore, it introduces control signals from the two modalities via feature selection. Only one modality guides music generation at any given time, leading to inadequate fusion between the two. V2Meow incorporates limited textual information, offering minimal guidance for music generation.
The unified integration of multiple modalities remains a critical trend in multi-modal research.

\textbf{MuMu-LLaMA}~\cite{liu2024m2ugen} is a comprehensive visual-music understanding and generation model capable of generating music guided by images, videos and text. It employs ViT~\cite{dosovitskiy2021vit} for image understanding and ViViT~\cite{arnab2021vivit} for video understanding. LLaMA 2~\cite{touvron2023llama} acts as a bridge for multi-modal understanding, trained to process and integrate visual, textual and audio information. During music generation, visual encodings from the understanding modules are fed into LLaMA 2, and the final layer of LLaMA 2 is passed to an output projector. This projector converts the outputs of LLaMA 2 into a music description embedding vector, which serves as a conditioning signal for pre-trained music generation models like AudioLDM~\cite{liu2023audioldm,liu2024audioldm2} and MusicGen~\cite{copet2024musicgen}. Finally, the pre-trained model synthesizes the corresponding music based on the embedding vector.

\textbf{XMusic}~\cite{xmusic2025} is a novel framework for controllable symbolic music generation. It supports flexible multi-modal prompts including images, video, text, tags and humming. 
Parsing and extracting multi-modal information is one of the key challenges in existing approaches. XMusic addresses the challenge through two key components. XProjector maps multi-modal prompts into symbolic music elements, considering aspects like emotion, genre, and rhythm.  XComposer, which comprises a generator and a selector, produces emotionally controllable music. The generator utilizes an innovative symbolic representation to create music, while the selector employs a multi-task learning approach to evaluate quality and musical attributes.

\section{Multi-Modal Music Datasets}

\begin{table}
\centering
\caption{Multi-Modal Music Datasets}
\label{tab:multimodal_datasets}
\resizebox{\textwidth}{!}{%
\begin{tabular}{lllrl}
\toprule
\textbf{Dataset} & \textbf{Modalities} & \textbf{Genre} & \textbf{Size (files)} & \textbf{Access} \\
\midrule
\multicolumn{5}{c}{\textit{Score-Audio Datasets}} \\
\midrule
LMD~\cite{raffel2016lmd} & MIDI, Audio & Mixed & 170,000+ & \url{https://colinraffel.com/projects/lmd/} \\
MusicNet~\cite{thickstun2017musicnet} & MIDI, Audio & Classical & 330 & \url{https://zenodo.org/records/5120004\#.YpbPI-5ByUl} \\
MAESTRO~\cite{hawthorne2019maestro} & MIDI, Audio & Classical (Piano) & 1,184 & \url{https://magenta.tensorflow.org/datasets/maestro} \\
ASAP~\cite{foscarin2020asap} & MusicXML, MIDI, Audio & Classical (Piano) & 222 & \url{https://github.com/fosfrancesco/asap-dataset} \\
Slakh2100~\cite{manilow2019slakh2100} & MIDI, Audio & Mixed & 2,100 & \url{http://www.slakh.com/} \\
POP909~\cite{wang2020pop909} & MIDI, Audio & Pop & 909 & \url{https://github.com/music-x-lab/POP909-Dataset} \\
EMOPIA~\cite{hung2021emopia} & MIDI, Audio, Emotion & Pop (Piano) & 1,087 & \url{https://zenodo.org/records/5090631} \\
\midrule
\multicolumn{5}{c}{\textit{Text-Music Datasets}} \\
\midrule
AudioPairBank~\cite{sager2018audiopairbank} & Text tags, Audio & Mixed & 33,000+ & \url{http://audiopairbank.dfki.de/} \\
AudioSet~\cite{gemmeke2017audioset} & Text tags, Audio & Mixed & 1,789,621 & \url{https://research.google.com/audioset/} \\
MusicCaps~\cite{agostinelli2023musiclm} & Text descriptions, Audio & Mixed & 5,521 & \url{https://www.kaggle.com/datasets/googleai/musiccaps} \\
MidiCaps~\cite{Melechovsky2024midicaps} & Text descriptions, MIDI & Mixed & 168,407 & \url{https://huggingface.co/datasets/amaai-lab/MidiCaps} \\
Lee et al.~\cite{lee2019icomposer} & Lyrics, Melody (MIDI) & Chinese Pop & 1,000 & \url{https://github.com/hhpslily/iComposer} \\
Yu et al.~\cite{yu2021lstmgan} & Lyrics, Melody (MIDI) & English Pop & 12,197 & \url{https://github.com/yy1lab/Lyrics-Conditioned-Neural-Melody-Generation} \\
\midrule
\multicolumn{5}{c}{\textit{Visual-Music Datasets}} \\
\midrule
MUSIC~\cite{zhao2018sound} & Video, Audio & Instrument Performance & 714 & \url{https://github.com/roudimit/MUSIC_dataset} \\
URMP~\cite{li2018urmp} & Video, Audio, MIDI, Score & Classical Performance & 44 & \url{https://labsites.rochester.edu/air/projects/URMP.html} \\
Solos~\cite{montesinos2020solos} & Video, Audio, Skeletal data & Instrument Performance & 755 & \url{https://juanmontesinos.com/Solos/} \\
AIST~\cite{tsuchida2019aist} & Dance video, Music & Street Dance & 13,939 & \url{https://aistdancedb.ongaaccel.jp/} \\
AIST++~\cite{li2021aist++} & 3D dance, Music & Street Dance & 1,408 & \url{https://google.github.io/aichoreographer/} \\
LORIS~\cite{yu2023loris} & Video, Music & Sports + Dance & 10,375 & \url{https://huggingface.co/datasets/OpenGVLab/LORIS} \\
TikTok~\cite{zhu2022d2mgan} & Dance video, Music & Dance & 445 & \url{https://github.com/L-YeZhu/D2M-GAN} \\
HIMV-200K~\cite{HIMV200K} & Video, Music & MVs & 200,500 & \url{https://github.com/csehong/VM-NET} \\
SymMV~\cite{zhuo2023video} & Video, Audio, MIDI & Pop (Piano) & 1,140 & \url{https://github.com/zhuole1025/SymMV} \\
\midrule
\multicolumn{5}{c}{\textit{Comprehensive Datasets}} \\
\midrule
MuChin~\cite{wang2024muchin} & Audio, MIDI, Text & Mixed (Chinese) & 1,000 & \url{https://github.com/CarlWangChina/MuChin} \\
MelBench~\cite{melfusion} & Audio, Text, Image & Mixed & 11,250 & \url{https://schowdhury671.github.io/melfusion_cvpr2024/} \\
BGM909~\cite{li2024diffbgm} & Audio, Text, Video & Pop & 909 & \url{https://github.com/sizhelee/Diff-BGM} \\
Popular Hooks~\cite{wu2024popularhooks} & Audio, MIDI, Lyrics, Video & Pop & 38,694 & ~\url{https://huggingface.co/datasets/NEXTLab-ZJU/popular-hook} \\
\bottomrule
\end{tabular}%
}
\end{table}

High-quality multi-modal datasets are fundamental for the training and evaluation of multi-modal music generation models. Existing datasets primarily encompass modal combinations such as score-audio, text-music, and visual-music, yet they still face challenges such as limited scale and insufficient annotation granularity. Table~\ref{tab:multimodal_datasets} lists the representative datasets. This section will categorically introduce these datasets, outlining their construction methods, relevant applications, and inherent limitations. We will also discuss solutions to alleviate the scarcity of data.

\subsection{Score-Audio Datasets}

Score-audio datasets aim to provide models with aligned information between musical scores and audio. To achieve alignment between scores and audio, such datasets not only need to provide score-audio data pairs but also ensure their alignment in the time dimension.

The Million Song Dataset (MSD)~\cite{bertin-mahieux2011million} is a large-scale music dataset containing metadata and audio features of 1 million songs, offering metadata extracted via The Echo Nest API~\footnote{\url{http://developer. echonest.com}}. However, MSD only includes music audio and does not contain corresponding scores. Based on MSD, Raffel built the \textbf{Lakh MIDI Dataset} (\textbf{LMD})~\cite{raffel2016lmd}, which contains over 170,000 MIDI files. By matching with the audio data in MSD, LMD provides large-scale score-audio aligned data.

\textbf{MusicNet}~\cite{thickstun2017musicnet} is a classical music dataset that provides MIDI-audio alignment, including 330 freely licensed recordings totaling 34 hours, covering 10 composers and 11 instruments. The Dynamic Time Warping (DTW) algorithm was employed to align recordings with MIDI-format scores. However, the audio and digital scores in MusicNet come from different sources, and their alignment is not entirely accurate. Addressing this issue, Hawthorne et al.~\cite{hawthorne2019maestro} collected data from the International Piano e-Competition to build \textbf{MAESTRO}. MAESTRO is a large-scale piano music dataset containing 172 hours of high-quality audio and MIDI data, covering 1,184 performances and 430 pieces. The audio and MIDI in the dataset were synchronously recorded by digital pianos during performances and further finely aligned with a precision of 3 milliseconds. However, MAESTRO does not provide aligned score annotations. To address this gap, Foscarin et al.~\cite{foscarin2020asap} built \textbf{ASAP}. ASAP is a dataset of western classical piano music containing 222 digital scores and 1,068 performances, providing finely corrected and annotated scores in MusicXML and MIDI formats, along with some performance audio. Through an automated annotation process, ASAP achieves finely aligned score annotations in terms of beats, measures, time signatures and key signatures.

Alignment 
can also be ensured by synthesizing audio from scores. \textbf{Slakh2100}~\cite{manilow2019slakh2100} is a high-quality synthetic dataset providing MIDI-audio alignment, containing 2,100 songs totaling 145 hours of mixed audio and corresponding individual instrument tracks. The dataset is rendered from LMD~\cite{raffel2016lmd} using professional virtual instruments.
However, the synthesized audio is not recorded from real performances, lacking in realism and expressiveness.

Although LMD provides large-scale aligned data, it lacks fine-grained annotations at the score level. Datasets like MusicNet, MAESTRO, and ASAP only include classical music, limiting their application in modern music generation tasks. \textbf{POP909}~\cite{wang2020pop909} is a high-quality dataset containing piano arrangements of 909 pop songs, providing MIDI and audio formats. POP909 provides three tracks for each song, including lead melody, secondary melody, and piano accompaniment, precisely aligned with the original audio. The dataset, created by professional musicians, ensures consistent and high-quality arrangements. It also provides detailed score annotations such as beats, chords, and key signatures. Hsiao et al.~\cite{hsiao2021compound} proposed a dataset containing 1,748 pop piano pieces with audio and scores, totaling approximately 108 hours. This dataset, through deep learning and rule-based automatic transcription and synchronization methods, offers rich symbolic representations and beat information. 
\textbf{EMOPIA}~\cite{hung2021emopia} is a emotion-annotated pop piano music dataset which contains 1,087 segments from 387 songs, covering both audio and MIDI data. EMOPIA provides segment-level emotion annotations, facilitating researchers in analyzing the dynamic changes in music emotions. 

\subsection{Text-Music Datasets}

Text-music datasets typically include description-music datasets and lyrics-melody datasets. Description-music datasets involve paired data of textual descriptions and corresponding music, either in symbolic or audio formats. 
Lyrics-melody datasets emphasize the alignment between lyrical content and melodic phrasing. This section explores the construction and characteristics of text-music datasets, highlighting their importance in advancing text-to-music generation research.

\subsubsection{Description-Music Datasets}

Early music datasets used text labels to annotate music.
\textbf{AudioPairBank}\cite{sager2018audiopairbank} is a dataset comprising over 33,000 audio files with refined tags. It was constructed by selecting tags from the online database Freesound\footnote{\url{https://freesound.org}} and downloading corresponding audio files.
Although AudioPairBank provides text labels for each audio file, these labels are weak and may be inaccurate, with some label combinations potentially being unreasonable. Addressing this issue, Gemmeke et al.~\cite{gemmeke2017audioset} built \textbf{AudioSet}, a large-scale audio dataset with precise labels. It contains 1,789,621 10-second audio clips totaling 4,971 hours, covering 632 audio event categories, with 56\% of the data being music. The data in this dataset comes from YouTube~\footnote{\url{https://www.youtube.com}} videos, annotated by human annotators, with the final labels determined by majority voting to ensure accuracy.

However, the labels provided in AudioPairBank and AudioSet are not complete sentences, offering limited information on the textual modality. 
\textbf{MusicCaps}~\cite{agostinelli2023musiclm} is a high-quality description-music dataset based on AudioSet, containing 5,521 samples. Each 10-second music clip in MusicCaps is sourced from AudioSet and accompanied by detailed descriptions and tags. These annotations are provided by professional musicians and describe aspects such as melody, rhythm, instruments, and emotions. The dataset also provides a genre-balanced subset of 1,000 samples, ensuring an even distribution of different music genres. 
However, MusicCaps is manually annotated by professional musicians, resulting in a limited size that is insufficient for large-scale model training. Additionally, MusicCaps only provides music audio without corresponding musical scores. To address these issues, Melechovsky et al.~\cite{Melechovsky2024midicaps} constructed the \textbf{MidiCaps} dataset. MidiCaps is a large-scale MIDI-caption dataset containing 168,407 MIDI files with detailed text descriptions. The dataset was built by extracting musical features from LMD~\cite{raffel2016lmd} and generating text descriptions using Claude 3~\footnote{\url{https://www.anthropic.com/claude}}. 
Furthermore, \textbf{ShutterStock}~\footnote{\url{https://www.shutterstock.com/music}} and \textbf{Pond5}~\footnote{\url{https://www.pond5.com}} are large-scale music data websites that provide corresponding text descriptions. They serve as partial sources of training data in MusicGen~\cite{copet2024musicgen}.

\subsubsection{Lyrics-Melody Datasets}

In addition to description-music datasets, there are also datasets that provide lyrics-melody pairs. 
Since each syllable in the lyrics may correspond to one or more notes, providing fine-grained alignment between lyrics and melody is crucial for such datasets. Lee et al.~\cite{lee2019icomposer} introduced a lyrics-melody dataset for Chinese pop music, which contains 1000 MIDI files. 
The researchers extracted the lead melody and recruited 20 annotators with experience in instruments to align the lyrics with the notes. However, this dataset is relatively small in scale and limited in musical diversity. 
Yu et al.~\cite{yu2021lstmgan} constructed a larger-scale lyrics-melody dataset. This dataset was collected from LMD~\cite{raffel2016lmd} and Reddit~\footnote{\url{https://www.reddit.com/r/datasets/}}, containing 12,197 MIDI songs totaling 789.34 hours. Each song in the dataset includes alignment annotations of English lyrics and melodies at the syllable level. Nevertheless, 
there are still challenges related to imprecise temporal alignment, which remains one of the key issues in the field of lyrics-to-melody generation.

\subsection{Visual-Music Datasets}

Currently, there are datasets that align music with performance videos. 
The \textbf{MUSIC} dataset~\cite{zhao2018sound} is a music performance dataset collected from YouTube, containing synchronized audio and visual data across 11 distinct instrument categories. It includes 714 videos, featuring both solo and duet performances. The \textbf{URMP} dataset~\cite{li2018urmp}, in addition to music performance videos, offers scores and fine-grained note annotations. URMP contains 44 classical chamber music pieces, ranging from duets to quintets, totaling 1.3 hours. The dataset includes scores in MIDI and PDF formats, as well as individual and mixed audio tracks. Additionally, it provides combined performance videos and comprehensive annotations at both frame and note levels. 
URMP's data collection relies on recruiting professional musicians for recordings. This approach, however, limits the dataset's size, hindering its ability to support the training of large-scale models. The \textbf{Solos} dataset~\cite{montesinos2020solos} is a 
larger-scale music performance dataset sourced from YouTube. It includes 755 videos of solo music performance, covering 13 different instruments. The dataset also provides body and hand skeletal data estimated by OpenPose~\cite{cao2020openpose}, along with precise timestamps.

Music is not only present in performance videos but also widely found in activities such as dance and figure skating. The \textbf{AIST} Dance Database~\cite{tsuchida2019aist} contains 13,939 dance videos and 60 original dance music tracks, covering 10 street dance styles. Each style is accompanied by six music tracks of varying tempos, all composed by professional musicians. Professional dancers performed to these tracks, ensuring synchronization between the dance movements and the music. Each dance in AIST includes multi-view videos, covering basic, advanced, group, and special scenario performances, providing rich resources for studying the relationship between dance and music. 
Extending AIST with 3D information, Li et al.~\cite{li2021aist++} presented \textbf{AIST++}. They recovered the camera calibration parameters and the 3D human motions from the multi-view dance videos in AIST. AIST++ offers multi-view synchronized images and detailed annotations, including camera parameters, joint positions, and SMPL parameters~\cite{loper2015smpl}. It contains 5.2 hours of 3D dance videos and 1,408 dance sequences, covering 10 dance styles. However, AIST only includes dance videos, and the video lengths vary, with some being relatively short. 
Therefore, Yu et al.~\cite{yu2023loris} presented the \textbf{LORIS} dataset, extending AIST++ by selecting videos longer than 25 seconds. LORIS also incorporates videos and the corresponding background music from figure skating~\cite{FS1000,xu2019FisV} and floor exercise~\cite{shao2020finegym}. 
As a result, LORIS is an 86.43-hour video-music dataset, with all clips exceeding 25 seconds in length, significantly expanding the available data in this field. The \textbf{TikTok} dataset~\cite{zhu2022d2mgan} is a challenging benchmark for real-world scenarios. It includes 445 real-world dance videos, averaging about 12.5 seconds per video, covering 85 different songs. Compared to AIST and its extensions, TikTok's videos feature more diverse backgrounds and often include incomplete human skeletal data. This dataset offers a more complex and diverse data environment to enhance the performance of multi-modal models in real-world scenarios.

Beyond performance and sports, the real world contains even richer videos. Video-to-music generation requires more generalized video-music datasets. In reality, platforms like YouTube and TikTok~\footnote{\url{https://www.tiktok.com}} host large amounts of video data. However, many of these data are of low quality, presenting a considerable challenge for data cleaning. A potential solution to address this issue is to filter for Music Videos (MVs), which typically exhibit strong video-music consistency and generally maintain high quality. Hong et al.~\cite{HIMV200K} filtered MVs from the YouTube-8M dataset~\cite{abu2016youtube}, resulting in \textbf{HIMV-200K}. HIMV-200K contains 200,500 video-music pairs, including official MVs, parody MVs, and user-generated MVs, offering strong diversity and complexity. However, HIMV-200K lacks fine-grained annotations for musical features and still includes many low-quality data points. To overcome these issues, Zhuo et al.~\cite{zhuo2023video} constructed the \textbf{SymMV} dataset, which contains 1,140 video-audio pairs totaling 76.5 hours. The researchers first retrieved professional pianist performance videos from YouTube and then searched for the corresponding official MVs based on the metadata. The piano audio clips were transcribed into MIDI files, 
enabling fine-grained musical annotations.

\subsection{Comprehensive Datasets}
\label{sec:MultimodalDatasets}

The aforementioned datasets provide aligned data between music and other modalities, but they only include a single guidance modality and lack information from additional modalities. 


\textbf{MuChin}~\cite{wang2024muchin} is a multi-modal Chinese colloquial music description dataset containing 1,000 high-quality entries, covering audio, symbolic and text modalities. The researchers recruited both professionals and amateurs to annotate music data across multiple dimensions, providing comprehensive insights into music emotions, styles, structures, and more.
With its precise and diverse data, MuChin serves as a valuable resource for evaluating and fine-tuning multi-modal models in music-related tasks.

Chowdhury et al.~\cite{melfusion} introduced the \textbf{MelBench} dataset, which provides image, text, and audio modalities. MelBench is a large-scale multi-modal music dataset, containing 11,250 samples across 15 music genres. During the construction of MelBench, 18 professional musicians first selected 10-second music clips from YouTube videos. Then they provided detailed text descriptions and carefully selected corresponding images from the videos, creating highly aligned multi-modal data. 
Furthermore, the researchers also extended MusicCaps~\cite{agostinelli2023musiclm} into a multi-modal dataset containing images, text, and music audio by extracting images from corresponding YouTube videos or the web. 

Li et al.~\cite{li2024diffbgm} introduced the \textbf{BGM909} dataset, a high-quality multi-modal resource derived from POP909~\cite{wang2020pop909}. BGM909 comprises 909 pop music videos with corresponding text descriptions and detailed annotations. The dataset was curated by searching for official videos on YouTube and manually aligning them with the music files. Additionally, the researchers provided comprehensive annotations, including camera shots and stylistic elements, to enhance its utility for multi-modal music generation research.

\textbf{Popular Hooks}~\cite{wu2024popularhooks} is a large-scale multimodal music dataset comprising 38,694 musical hooks with synchronized MIDI, MVs, audio, and lyrics, along with rich annotations including tonality, structure, genre, region, and emotion. The dataset addresses the scarcity of aligned multi-modal music data by offering precise time alignment between modalities and comprehensive musical attributes, with emotion labels automatically generated using a pretrained multi-modal framework and validated through a user study. 

Despite the emergence of multi-modal datasets, existing datasets remain relatively small in scale and lack comprehensive modality coverage. The rapid development of multi-modal models currently demands a large-scale, diverse, and comprehensive dataset that encompasses text, scores, audio, emotions, visuals, and more. The exploration of multi-modal datasets in the field of music generation is still in its early stages.

\subsection{Solutions for Multi-Modal Data Scarcity}

As mentioned above, 
existing multi-modal music datasets struggle to meet the demands of large-scale models. 
Therefore, how to fully utilize large-scale single-modal data to train multi-modal models is one of the key points considered by many multi-modal music generation methods.

Some approaches use pre-trained models to automatically add matching multi-modal information to single-modal data, thereby obtaining aligned multi-modal data. 
Liu et al.~\cite{liu2024m2ugen} proposed an automated multi-modal data collection scheme based on pre-trained multi-modal models. They first obtained music samples from AudioSet~\cite{gemmeke2017audioset} and retrieved the corresponding videos based on the metadata. Then they randomly extracted frames from the videos as the image modality. Subsequently, they used specialized multi-modal models~\cite{liu2024mullama,li2022blip,tong2022videomae,MosaicML2023mpt} to generate captions and instructions as the text modality.
Additionally, MuseCoco~\cite{lu2023musecoco} and Mustango~\cite{melechovsky2024mustango} use ChatGPT~\footnote{\url{https://openai.com/chatgpt/overview}} to transform label-based musical features and metadata into textual descriptions. Noise2Music~\cite{huang2023noise2music} uses LLMs to generate a series of music text descriptions, which are then matched with music audio using MuLan~\cite{mulan} to construct text-music paired data.

Other methods supplement single-modal data with multi-modal information by crawling data from the web. Moûsai~\cite{mousai-2024-efficient} uses corresponding tags from music websites as textual descriptions, while ERNIE-Music~\cite{zhu2023erniemusic} utilizes music comments from these sites as music descriptions, building datasets based on this content.

Enhancing model architectures or training methods can enable models to learn from single-modal data, reducing the need for highly aligned multi-modal data. SongMASS~\cite{sheng2021songmass} introduces pre-training methods to allow models to learn from unpaired single-modal data, using only a small amount of paired data to train the model's cross-modal transfer capabilities. TeleMelody~\cite{ju2022telemelody} introduces intermediate representations for two modalities, splitting cross-modal tasks so that model training no longer relies on paired data. CMT~\cite{di2021cmt} introduces a rule-based mapping relationship between video rhythm and music. Following CMT, XMusic~\cite{xmusic2025} utilizes mapping algorithms and pre-training understanding models to map various modalities to musical features. The two models are both trained on single-modal datasets, but they can be applied to cross-modal and multi-modal music generation through rule-based mapping methods.


\section{Evaluation}

Currently, there exist a number of multi-modal music generation algorithms, making the evaluation of these methods a significant concern. However, music is an art that is constrained by objective factors such as acoustics and music theory, while also being influenced by subjective factors like personal experiences, expertise, and aesthetics. Thus, assessing the quality of music is not straightforward. It requires both an objective evaluation based on theoretical grounds and a subjective evaluation from human listeners. Moreover, multi-modal music generation emphasizes the use of information from the guidance modalities to produce music that meets specific requirements. Consequently, the evaluation of multi-modal music generation extends beyond the assessment of inherent musical quality. A crucial dimension is the evaluation of cross-modal consistency, which quantifies the fidelity with which the generated music reflects the input modalities. 
These aspects pose challenges to evaluation methods, necessitating the design of robust evaluation systems to assess multi-modal music generation approaches. Although there is no unified evaluation system in the field of multi-modal music generation, researchers are striving to propose more reasonable and efficient evaluation metrics. This section will introduce the widely used subjective and objective metrics for multi-modal music generation.

\subsection{Objective Metrics}

Although music is an art with strong subjective elements, thanks to the theoretical research, there are now some methods that can objectively evaluate the overall aspects and certain dimensions of music. 
These methods typically use mathematical formulas to calculate the similarity between generated music and real music or music theory, thereby assessing the rationality and quality of the generated music. In terms of modal consistency evaluation, objective consistency evaluation methods calculate the semantic similarity between the generated music and the guidance modalities, or the faithfulness of the generated music to the control conditions, to reflect the consistency between the generated music and the guidance modalities. However, it is important to note that music objective evaluation methods can only reflect the similarity between the generated music and the ground truth, but not directly indicate the quality of the generated music in human perception. Moreover, music is complexly composed of various elements and aspects. Therefore, objective evaluation methods often cannot comprehensively assess all aspects of music.

\subsubsection{Audio}

The evaluation of music audio primarily focuses on musical attributes such as quality and structure, as well as aspects like originality and diversity.

\textbf{Quality.} \textbf{Fréchet Inception Distance (FID)}~\cite{FID} is an objective evaluation method based on the Fréchet distance~\cite{frechet1957distance}, commonly used in the field of image generation to measure the distribution difference between generated and real images in the latent space. Yang et al.~\cite{yang2023diffsound} adapted the InceptionV3 model~\cite{szegedy2016inceptionv3} used in the original FID metric and trained it on AudioSet~\cite{gemmeke2017audioset} to make it suitable for audio. This modified FID can capture the similarity of generated audio to real audio. It calculates the mean and covariance matrix of the generated and real audio in the modified InceptionV3 feature space and further computes the distance between them. A low FID value indicates that the generated audio is close to the real audio in distribution, implying high generation quality.

Similarly based on the Fréchet distance, \textbf{Fréchet Audio Distance (FAD)}~\cite{roblek2019FAD} uses the audio classification model VGGish ~\cite{hershey2017vggish} as the audio understanding model, serving as a no-reference metric to evaluate the quality of generated music. It aims to measure the perceptual difference between generated music and studio-quality music. Unlike traditional signal-level metrics, FAD does not require the original clean music as a reference. Instead, it assesses quality by comparing the embedding statistics of generated music with those of a set of clean music. Specifically, FAD uses VGGish to extract embeddings from audio. It then calculates the Fréchet distance between the multivariate Gaussian distributions of the generated music and the background statistics computed on a large set of studio recorded music. A low FAD value indicates that the generated music is close to studio-quality music.


\textbf{Structure.} The evaluation of long-term structure aims to analyze whether the generated music possesses a reasonable segmented structure. 
Schneider et al.~\cite{mousai-2024-efficient} proposed a method for assessing the long-term structure of music by analyzing changes in the average amplitude and variance across different musical segments. A well-structured piece of music typically follows a gradual introduction, a main body with the core content, and a gradual conclusion. Thus, the average amplitude and variance of its segments should reflect this progression. 
However, relying solely on changes in amplitude and variance may not fully reflect the complexity of musical structures. Therefore, this evaluation method serves more as an auxiliary tool for preliminary judgments on the structural rationality of music.

\textbf{Originality.} LLMs have the capacity to memorize patterns seen in the training data~\cite{carlini2021extracting}. Models might achieve high scores by excessively memorizing the training data and producing content that is overly similar to it. However, generating content that is too similar to the training data is a phenomenon that music generation models should avoid. For this issue, Agostinelli et al.~\cite{agostinelli2023musiclm} introduced a training data memorization analysis method~\cite{carlini2023quantifying} to assess whether the model 
generates content that is too similar to the training data. By comparing the semantic tokens of the generated content with those in the training data, the method calculates the proportions of exact matches and approximate matches. A lower degree of memorization indicates that the generated audio is less similar to the training data. 

\textbf{Diversity.} \textbf{Inception Score (IS)}~\cite{salimans2016inceptionscore} is a metric used to assess the clarity and diversity of generated samples. Kong et al.~\cite{kong2021diffwave} introduced this metric into the field of audio generation. 
Specifically, IS first uses a pre-trained classification model 
to classify the generated samples, obtaining the class probability distribution for each sample. Then, IS computes the Kullback-Leibler Divergence (KLD) between the class probability distribution of the generated samples and their marginal class distribution, converting the result into an easily interpretable score through an exponential function. A higher IS value indicates better class clarity and diversity of the generated samples. While IS can assess inter-class diversity of generated samples, it does not consider within-class diversity. \textbf{Modified Inception Score (mIS)}~\cite{gurumurthy2017minceptionscore} is an extension of IS, designed to evaluate the within-class diversity of generated samples. Similar to IS, mIS also uses a pre-trained classification model to classify the generated samples, obtaining the class probability distribution for each sample. Then, mIS computes the KLD between the class probability distributions of the generated samples and converts the result through an exponential function. A higher mIS value indicates better within-class diversity. 

\subsubsection{Symbolic Music}

Unlike audio, symbolic music contains rich note information, making it easy to extract detailed musical features and conduct evaluations based on them. Therefore, compared to music audio, objective evaluation methods for symbolic music delve deeper into musical details and are more diverse. Currently, researchers have proposed a multitude of evaluation methods targeting various aspects of symbolic music.

Due to the ease of extracting musical features from symbolic music, existing evaluation methods typically conduct a multi-dimensional assessment of melody, pitch, rhythm, and structure, in addition to an overall evaluation of the music, to comprehensively assess its quality. For example, Yang and Lerch~\cite{yang2020evaluation} proposed a set of objective evaluation metrics for symbolic music, reflecting the quality of generated music by comparing the data distribution of generated music with that of the training data based on both absolute and relative metrics. Initially, this method extracts features related to pitch and rhythm, such as pitch count, pitch class histogram, and note count. 
Then, on the one hand, the absolute metrics calculate the mean and standard deviation of the features to describe the overall characteristics. On the other hand, relative metrics generate probability density functions (PDFs) through pairwise cross-validation and kernel density estimation, and then compute the \textbf{KLD} and \textbf{Overlap Area (OA)} to measure the distribution similarities between the generated music and the training dataset.
The researchers also open-sourced the mgeval toolkit~\footnote{\url{https://github.com/RichardYang40148/mgeval}}, which implements the above methods and provides visualization tools. This set of metrics has high reproducibility and reliability, offering valuable references for optimizing music generation systems. However, this method primarily measures the similarity between generated music and the training set and does not fully reflect the quality of the music.

Zhuo et al.~\cite{zhuo2023video} provided a set of evaluation metrics based on music theory. 
These metrics include \textbf{Scale Consistency (SC)}, \textbf{Pitch Entropy (PE)}, \textbf{Pitch Class Entropy (PCE)}, \textbf{Empty Beat Rate (EBR)}, and \textbf{Inter-Onset Interval (IOI)}. Each metric evaluates different aspects of the music, with values closer to theoretical ideals and that of real music indicating better quality. SC measures whether the pitch of the music conforms to a scale, assessing the harmony of the music by checking if the pitches follow specific scale rules. 
PE measures the diversity of pitches by calculating the entropy of pitch values. 
PCE measures the diversity of pitch classes, similar to PE but focusing on pitch class variation. 
For the evaluation of rhythm, EBR measures the proportion of empty beats and IOI measures the time intervals between notes. The closer these two indicator values are to the statistical values of real music, the more similar the generated music is to real music.
These metrics collectively provide a comprehensive evaluation of music, objectively measuring generated music from multiple dimensions. Through these metrics, researchers can better understand the characteristics of generated music and further optimize generation models.


In addition to the above methods, there are other metrics that evaluate different aspects of symbolic music.

\textbf{Structure.} \textbf{Similarity Error (SE)}~\cite{yu2022museformer} is a metric used to evaluate the structural rationality of the generated music. Specifically, SE reflects how closely the generated music resembles human-composed music in terms of structural features such as repetition and variation. A lower SE value indicates that the structure of the generated music is more similar to that of the training data. During the calculation process, SE is based on the similarity statistics of different time intervals in the music, effectively capturing both short-term and long-term structural features in the music. \textbf{Grooving Pattern Similarity (GS)}~\cite{wu2020jazz} provides an objective measure of rhythmic consistency within a musical piece. The grooving pattern represents the positions in a bar at which there is at least a note onset. This metric operates on 64-dimensional binary vectors representing the grooving patterns. By computing the XOR-based similarity between grooving patterns of different bars, GS quantifies the degree of rhythmic regularity. High GS values correspond to predictable, consistent rhythms. 
Structureness, a fundamental aspect of musical form, reflects the presence of repeating musical material in the composition at various granularities, from short motifs to entire sections. To quantify this, Wu et al.~\cite{wu2020jazz} proposed the \textbf{Structureness Indicator (SI)}. SI is derived from the fitness scape plot~\cite{muller2011segment,muller2012scape}, which visualizes repeating patterns within a musical piece. These indicators focus on identifying the most prominent repetitions within specific time intervals. By examining the structureness in various ranges, researchers can analyze the structural organization of a composition at multiple levels. An SI value close to that of the training data suggests that the generated music exhibits a similar degree of structural repetition as the real music.

\textbf{Tonality.} 
\textbf{Pitch Class Histogram Entropy}~\cite{wu2020jazz} quantifies the tonal clarity of generated music by measuring the uncertainty of its pitch distribution. Initially, a histogram is created to represent how often each of the 12 pitch classes appears. The entropy of this histogram is then calculated, measures the uncertainty of the pitch distribution. A low entropy value indicates that certain pitch classes dominate, suggesting a clear and stable tonality. 

\textbf{Melody.} Sheng et al.~\cite{sheng2021songmass} proposed a set of melody evaluation metrics. \textbf{Pitch Distribution Similarity (PD)} reflects whether the overall pitch distribution of the generated melody is reasonable. Specifically, PD calculates the pitch frequency histograms of both the generated and real melodies and compares their overlap area. 
A larger overlap area indicates that the generated melody's pitch distribution is closer to that of the real melody. 
\textbf{Duration Distribution Similarity (DD)} reflects the rationality of the rhythm and note durations. 
Similar to PD, DD calculates the duration frequency histograms of both the generated and real melodies and compares their overlap area. \textbf{Melody Distance (MD)} is used to assess the similarity in pitch trends between generated melodies and real melodies. This method unfolds the notes in a time series and calculates the difference between each note's pitch and the average pitch of the entire sequence. Then, the DTW algorithm is used to compute the distance between the generated and real melody sequences. A smaller MD indicates that the pitch trend of the generated melody is closer to that of the real melody.

\subsubsection{Consistency}

When evaluating modal consistency of music generation systems, one can extract music semantics using music understanding models and compare the semantic similarity between the generated music and the original data. In addition to using pre-trained music understanding models, contrastive embedding models such as CLAP~\cite{clap} and MuLan~\cite{mulan} can also be employed, where the similarity between the guidance modality and the music audio in the embedding space represents their consistency. However, these semantic consistency evaluation methods can only assess the consistency between the generated music and the guidance modalities at a macro level, such as emotion, atmosphere, and style. It is also necessary to evaluate the consistency in fine-grained musical elements, which pertains to the controllability of the model.

\textbf{Evaluation Based on Music Understanding Models.} \textbf{Kullback-Leibler Divergence (KLD)} is an objective evaluation metric used to measure the distribution difference in semantic labels between generated music and reference music. It first obtains the label distributions of the generated music and reference music using a pre-trained classification model, and then calculates the Kullback-Leibler divergence between the label distributions of the generated music and reference music. A lower value indicates that the generated music is more semantically similar to the reference music. \textbf{Audio Caption Loss}~\cite{yang2023diffsound} is a metric used to measure the relevance between generated audio and textual descriptions. It inputs the generated audio into an audio captioning model to produce a textual description, and then calculates the difference between the generated text and the original text. However, these metrics rely on pre-trained classification models or captioning models, and thus their performance is limited by the quality of the pre-trained models. Additionally, these methods can only capture semantic differences between generated music and reference music, and cannot reflect other characteristics of the music.

\textbf{Evaluation Based on Contrastive Embedding Models.} \textbf{CLAP Score}~\cite{clap} assesses text-audio alignment by calculating the cosine similarity between the generated music and the textual description in the joint embedding space of CLAP. A higher score indicates a better match between the generated music and the textual description. Similarly, \textbf{MuLan Cycle Consistency (MCC)}~\cite{agostinelli2023musiclm} measures the semantic consistency by computing the cosine similarity in the MuLan embedding space~\cite{mulan}. 
For evaluating the semantic consistency between generated music and input images, Chowdhury et al.~\cite{melfusion} proposed a new metric, \textbf{Image Music Similarity Metric (IMSM)}. IMSM combines the text-image embedding model CLIP~\cite{radford2021clip} and the text-audio embedding model CLAP~\cite{clap}, using text as a bridge to connect image and audio semantics. 
Similarly, 
Liu et al.~\cite{liu2024m2ugen} introduced \textbf{ImageBind Ranking (IB Rank)}, using the ImageBind model~\cite{girdhar2023imagebind} to measure the relevance between generated music and the input images or videos. ImageBind is a multi-modal alignment model capable of mapping different modalities into the same embedding space, enabling cross-modal semantic alignment. In IB Rank, the embeddings of the input and the generated music are first extracted using ImageBind, and then the cosine similarity between these embeddings is calculated.
However, it is important to note that such evaluation methods relying on joint embedding models depend on the quality of the pre-trained models. Furthermore, although these methods can directly measure semantic alignment, they cannot reflect other characteristics of the music.

\textbf{Controllability Evaluation.} Controllability refers to the ability of a music generation model to precisely control specific musical attributes in the output music based on guidance modalities. 

In the evaluation of controllability in text-to-music generation, Melechovsky et al.~\cite{melechovsky2024mustango} proposed a series of controllability metrics to measure the accuracy of generated music in specific musical attributes. \textbf{Tempo Bin (TB)} and \textbf{Tempo Bin with Tolerance (TBT)} are used to assess whether the tempo of the generated music matches the true tempo. TB requires the tempo of the generated music to fall within the true tempo range, while TBT allows it to fall within adjacent ranges. \textbf{Correct Key (CK)} and \textbf{Correct Key with Duplicates (CKD)} evaluate whether the key of the generated music matches the true key. CK requires the key of the generated music to exactly match the true key, while CKD allows it to match relative keys. \textbf{Perfect Chord Match (PCM)} and \textbf{Exact Chord Match (ECM)} assess whether the chord sequence of the generated music matches the true chord sequence. PCM requires the chord sequence of the generated music to fully match the true sequence, while ECM allows for missing or extra chord instances. \textbf{Chord Match in any Order (CMO)} and \textbf{Chord Match in any Order major/minor Type (CMOT)} evaluate the proportion of chords in the generated music that match the true chord sequence. CMO does not require the order to match, while CMOT further requires matching chords to have the same root and chord type (major/minor). \textbf{Beat Match (BM)} assesses whether the beat count of the generated music matches the true beat count. These metrics quantify the accuracy of generated music in specific musical attributes, helping to evaluate the controllability of the model. However, they rely on the accuracy of music feature extraction and may not be sensitive enough for complex musical structures.

In the evaluation of controllability in video-to-music generation, Yu et al.~\cite{yu2023loris} improved upon the metrics \textbf{Beats Coverage Scores (BCS)} and \textbf{Beat Hit Scores (BHS)} from music-to-motion generation methods~\cite{lee2019dancing,davis2018visual}, proposing a series of metrics to assess the beat synchronization. Specifically, BCS measures the proportion of the beats in the reference music covered by the beats in the generated music, and BHS measures the proportion of the beats in the reference music hit by the beats. Additionally, the standard deviations of BCS and BHS are calculated to measure the generative stability, termed as \textbf{Coverage Standard Deviation (CSD)} and \textbf{Hit Standard Deviation (HSD)}, respectively. The F1 scores of BCS and BHS are also calculated as an integrated assessment.

\subsection{Subjective Metrics}


As current objective methodologies fail to comprehensively assess all aspects of musical perception and experience, subjective music evaluation is indispensable. Here are several common subjective evaluation methods.


\begin{description}
\item[Rating] Participants assign numerical scores to indicate their perceived quality or other attributes of the music.
\item[Preference] Participants compare music samples and indicate which one they prefer based on specific criteria.
\item[Turing Test] Participants listen to music samples and try to identify which samples are human-composed and which are AI-generated.  The AI's ability to be misidentified as human is a measure of its quality.
\item[Forced Choice] Participants hear a piece of music, then answer questions each with multiple choices, with only one being the target. The proportion of correct answers indicates the prompt fidelity of the generative model.
\end{description}


In the evaluation of multi-modal music generation, subjective evaluation methods primarily assess two aspects: music quality and modal consistency. 
This section will provide detailed examples of how the aforementioned subjective evaluation methods are applied to assess music quality and modal consistency.

\subsubsection{Musical Quality}
\label{sec:SubEvlMusQual}

\textbf{Humans Mean Opinion Score (MOS)}~\cite{yang2023diffsound} is a subjective evaluation metric that directly reflects human perception of audio quality. The participants rate the audio based on three aspects including relevance, fidelity, and intelligibility. In practice, the researchers randomly selected 15 sets of generated audio samples, each including a text description, a real sample, 1-2 audio samples generated by the baseline model, and 1-2 audio samples generated by the proposed model. Then, 10 participants were asked to rate each audio sample, and the average score was taken as the MOS rating.
Building on the experimental setup of MOS, Kreuk et al.~\cite{kreuk2023audiogen} proposed the \textbf{Overall Quality (OVL)} metric to measure the overall perceived quality of generated music. Participants score the overall quality of the generated music on a scale from 1 to 100, with higher values indicating better quality. OVL can capture subtle differences in audio, especially in terms of perceived quality.

Melechovsky et al.~\cite{melechovsky2024mustango} introduced a set of subjective evaluation methods in MusTango to assess the overall quality of generated music, its relevance to text prompts, and the controllability of specific musical attributes through subjective ratings. 
To reduce subjective bias from differing musical experience and offer a comprehensive evaluation of the music, the researchers recruited both general listeners and professional musicians as participants. The overall evaluation of the music audio includes two dimensions, \textbf{Audio Rendering Quality (AQ)} and \textbf{Overall Music Quality (OMQ)}, which evaluate the sound quality and musicality of the audio, respectively.

The subjective evaluation method proposed by Schneider et al.~\cite{mousai-2024-efficient} covers multiple dimensions, including the authenticity, melodiousness, harmony, and audio clarity of the music. The Music Turing Test requires participants to listen to a pair of music samples, identify which one is generated, and rate the similarity of the generated music to the real music. If the participants correctly identifies the generated music, their rating is recorded. Conversely, if they incorrectly identify it, the generated music is automatically given a score of 5, indicating it has fully passed the Turing Test. Next, the Melodiousness Evaluation requires participants to rate the beauty of the melody based on its rhythm and repetitiveness. The Harmony Evaluation requires participants to assess whether multiple notes are harmonious and supportive of the main melody. Finally, the Audio Clarity Evaluation requires participants to judge the clarity of the audio, transitioning from a walkie-talkie to high-quality studio audio.

\subsubsection{Consistency}

\textbf{Relevance (REL)}~\cite{kreuk2023audiogen} is a subjective evaluation metric that uses human ratings to measure the degree of match between the generated music and the guidance modality. Participants score the alignment between the generated music and the guidance modality on a scale from 1 to 100, with higher values indicating a better match.

In the subjective evaluation method of MusTango mentioned in Section~\ref{sec:SubEvlMusQual}, the general listeners are required to assess the overall quality of the generated music, its relevance to text prompts, and its musicality, among other aspects. In addition to the overall metrics, participants also rate the generated music and text prompt pairs on dimensions such as \textbf{Relevance to Text Prompt (REL)}, \textbf{Rhythm Consistency (RC)}, and \textbf{Harmony and Consonance (HC)}. 
The experts rate dimensions like \textbf{Chord Match (MCM)} and \textbf{Tempo Match (MTM)} based on the generated music and text prompts. 
Combining general listener testing and expert listener testing allows for a more accurate and comprehensive assessment of the overall quality and controllability of the generated music.

The human subjective evaluation used in MusicLM~\cite{agostinelli2023musiclm} adopts an A vs B comparison approach. Listeners are asked to compare two audio clips and select which one better matches the text description. 
The final results tally the number of "wins" for each model, with higher scores indicating a better match. In this subjective evaluation, the researchers normalized the audio quality to eliminate its impact on the subjective assessment.

Schneider et al.~\cite{mousai-2024-efficient} employed text-music relevance evaluation using a four-alternative forced-choice (4AFC) paradigm, requiring annotators to infer the genre of the generated music. By calculating the proportion of correct classifications, the method evaluates whether the generated music accurately reflects the content of the text prompts. 

\section{Challenges and Future Directions}

Although significant progress has been made in multi-modal music generation, the field still confronts core challenges such as modal fusion, data scarcity and an imperfect evaluation system. This section will delve into the current technical bottlenecks and propose future research directions, providing a systematic reference for subsequent studies.

\subsection{Challenges}

To our best knowledge, the multi-modal music generation may still face the following challenges.

\textbf{Creativity.} Existing models often rely heavily on training data, resulting in lack of creativity. Exploring how to introduce more creative elements into models to generate unique music is a direction that requires further investigation.

\textbf{Efficiency.} The trend in generative models is toward increasing the model scale. However, music, as sequential data, typically adopts autoregressive generation methods, which some studies~\cite{copet2024musicgen,lam2024MeLoDy} have found to be inefficient. The efficiency of models impacts the development and application of music generation technologies, especially when incorporating information from more modalities. Ensuring high efficiency while maintaining generation quality is an issue that still requires further exploration.


\textbf{Modal Fusion.} Multi-modal music generation involves the fusion of information from various modalities. Inconsistencies or conflicts may arise between different modalities, and how to achieve effective fusion in such scenarios is a problem that requires in-depth research.

\textbf{Modal Consistency.} Ensuring consistency between generated music and the guidance modalities is a critical challenge. Modal differences make it difficult to comprehensively reflect the guidance modality. Moreover, modal consistency not only involves content alignment, but also extends to emotional, stylistic, and other multidimensional consistencies, necessitating more refined model design and evaluation methods.

\textbf{Data.} multi-modal music generation requires large amounts of multi-modal data for training. However, acquiring high-quality multi-modal data is challenging. Existing datasets are limited in scale and the number of modalities they include. Additionally, data annotation and quality control are significant issues, particularly in ensuring alignment and consistency across different modalities.

\textbf{Evaluation.} Both objective and subjective evaluation methods have difficulties in comprehensively evaluating music. There is a need for a unified evaluation system to assess the continuously emerging music generation models. Furthermore, multi-modal music generation emphasizes modal alignment, requiring a set of methods to accurately and comprehensively evaluate the consistency, ensuring that the generated music faithfully reflects the multidimensional information provided by the guidance modalities.

\textbf{Application.} Although a variety of multi-modal music generation technologies have emerged and been applied, several issues currently limit their further application. Firstly, the sound quality of music generated by existing products still significantly lags behind professionally released music. Secondly, the professionalism, controllability and editability of existing music generation products are not strong, making the music they generate difficult to apply in more professional scenarios. Finally, the use threshold for existing music generation products is relatively high, requiring users to comprehensively describe the music to better meet their needs~\cite{wang2024muchin}.

\subsection{Future Directions}


To address the aforementioned challenges, the following future research directions remain to be explored:

\begin{itemize}
    \item By introducing multi-modal information to inspire the creativity of music generation models, generating creative music rather than merely imitating and piecing together training data.
    \item Improving generation efficiency while ensuring quality through comprehensive optimization at the algorithmic architecture and hardware-software levels, enabling real-time music generation.
    \item Comprehensively utilizing pretrained models from different fields and modalities, combining them into high-performance multi-modal music generation models.
    \item Ensuring that generated music is highly aligned with the guidance modalities, comprehensively and faithfully reflecting the input information across dimensions such as style, emotion, and semantics.
    \item Constructing large-scale, comprehensive, and multi-modal datasets with high alignment across modalities to support the training of large-scale multi-modal music models.
    \item Building a comprehensive, systematic, and multidimensional evaluation system for multi-modal music generation methods, capable of accurately assessing not only the quality of generated music but also its artistic value, creativity, and modal consistency.
    \item Enhancing the quality and efficiency of music generation. Improving professionalism, controllability, and editability. Increasing user-friendliness and lowering the barrier to entry. Ultimately, achieving the full implementation of multi-modal music generation technology in both professional and everyday scenarios.
\end{itemize}


\section{Conclusion}

This paper systematically reviews the development trajectory of music generation technology from a modal perspective, summarizing the model architectures, datasets, and evaluation systems of single-modal, cross-modal, and multi-modal methods. Although existing technologies have shown potential in specific scenarios, issues such as lack of creativity and insufficient modal alignment still constrain their practical application. Future research needs to further explore the deep fusion mechanism of multi-modal information, construct generative frameworks with strong versatility and high controllability, and promote the transition of music generation technology from the laboratory to industrial application.

\bibliographystyle{ACM-Reference-Format}
\bibliography{sample-base}










\end{document}